\documentclass[a4paper,fleqn,usenatbib]{mnras}
\usepackage{aas_macros}
\usepackage{mathtools}
\usepackage{hyperref}
\usepackage{breqn}
\usepackage[T1]{fontenc}
\usepackage{ae,aecompl}
\usepackage{tablefootnote}

\usepackage{graphicx}	
\usepackage{amsmath}	
\usepackage{amssymb}	
\usepackage{subfig}
\usepackage{float}
\usepackage{multirow}
\usepackage{units}
\usepackage{placeins}
\usepackage{pdflscape}
\usepackage[table]{xcolor}
	
\title[Long-term evolution of BH-ULX candidates]{Long-term evolution of BH-ULX candidates: An `unusual' $L_{\rm disc} - T_{\rm col}$ correlation associated with spectral states}

\author[Majumder et al.]{Seshadri Majumder$^{1}$\thanks{E-mail: smajumder@iitg.ac.in},
	Santabrata Das$^{1}$\thanks{E-mail: sbdas@iitg.ac.in}
	and
	Anuj Nandi$^{2}$\thanks{E-mail: anuj@ursc.gov.in} \\
	$^{1}$Department of Physics, Indian Institute of Technology Guwahati, Guwahati, 781039, India.\\
	$^{2}$Space Astronomy Group, ISITE Campus, U. R. Rao Satellite Centre, Outer
	Ring Road, Marathahalli, Bangalore, 560037, India.
}

\date{Accepted XXX. Received YYY; in original form ZZZ}

\pubyear{2024}

\begin{document}
	\label{firstpage}
	\pagerange{\pageref{firstpage}--\pageref{lastpage}}
	\maketitle

\begin{abstract}

We present the long-term spectral evolution of eight black hole ultra-luminous X-ray sources (BH-ULXs), namely NGC$1313$ X$-1$, NGC$5408$ X$-1$, NGC$6946$ X$-1$, IC$342$ X$-1$, NGC$55$ ULX$1$, NGC$4395$ ULX$1$, NGC$5204$ X$-1$ and NGC$4190$ ULX$1$ using {\it XMM-Newton} monitoring data spanning over a decade or more.  An in-depth spectral modeling with thermal Comptonization (\texttt{nthComp}) and standard disc (\texttt{diskbb}) components reveals NGC$5204$ X$-1$, IC$342$ X$-1$, NGC$4190$ ULX$1$ and NGC$1313$ X$-1$ exhibiting harder spectral characteristics with dominant effect of Comptonization ($F_{\rm nth}>F_{\rm disc}$, $\Gamma_{\rm nth}\lesssim2$). However, NGC$6946$ X$-1$ and NGC$55$ ULX$1$ remain in a disc-dominated state ($F_{\rm disc}\sim2F_{\rm nth}$, $\Gamma_{\rm nth}\gtrsim 2$), while NGC$5408$ X$-1$ shows intermediate spectral characteristics. The spectral analyses indicate an anti-correlation between disc luminosity ($L_{\rm disc}$) and temperature ($T_{\rm col}$) for all sources except NGC$5204$ X$-1$. These anti-correlations follow a relation $L_{\rm disc} \propto T_{\rm col}^{\alpha}$ with steeper exponents of $\alpha=-6.01\pm0.25$ (NGC$55$ ULX$1$), $-8.93\pm0.11$ (NGC$6946$ X$-1$), and $-10.31\pm0.10$ (NGC$5408$ X$-1$) for sources with softer or intermediate spectral characteristics. For harder sources, NGC$1313$ X$-$1 and IC$342$ X$-$1, the combined results provide $\alpha=-3.58\pm0.04$. However, for NGC$5204$ X$-1$, a positive correlation is found, yielding $\alpha=1.4\pm0.1$, suggesting that the emission mechanism is associated with the transition from the `standard disc' to the `slim disc' scenario. These findings suggest that the observed $L_{\rm disc}-T_{\rm col}$ correlations, along with the overall spectro-temporal properties of BH-ULXs, seems to be governed by disc-corona-wind driven accretion processes at various inclinations. Finally, we report a QPO-like feature ($\sim20$ mHz) with $rms\%\sim 6.6$, Q-factor $\sim6.7$ and significant $2.8\sigma$ in NGC$55$ ULX$1$.

\end{abstract}


\section{Introduction}

Ultraluminous X-ray sources (ULXs) are the class of extra-galactic, point-like, non-nuclear objects with isotropic luminosity in excess of $10^{39}$ erg $\rm s^{-1}$ \citep{Fabbiano-etal1989, Makishima-etal2000}. Despite being discovered more than thirty years ago, the exact nature of these objects remains a point of debate and their observational features are still not well understood (see \citealt[for a recent review]{Feng-etal2011, Kaaret-etal2017, Fabrika-etal2021, King-etal2023}). Based on the observational evidences from timing and spectral variability, a population of the ULXs are believed to be powered by super-Eddington accretion onto stellar-mass black holes \cite[see references therein]{Bachetti-etal2013, Walton-etal2013, Walton-etal2014, Walton-etal2015a, Rana-etal2015}. The presence of massive stellar-mass black holes in several ULXs is also suggested \citep{Agrawal-etal2015, Das-etal2021, Majumder-etal2023}. Alternatively, a plausible scenario relying on the intermediate-mass black holes (IMBHs) of mass $\gtrsim 100 M_\odot$ as the powerhouse of such extremely luminous systems is also propounded \citep{Colbert-etal1999, Makishima-etal20,Miller-etal2003, Das-etal2021, Majumder-etal2023}. In addition, the possibility of having highly magnetized neutron stars at the central core is also put forward to explain the pulsations detected in several ULXs \citep{Bachetti-etal2014, Furst-etal2016, Israel-etal2017, Carpano-etal2018, Sathyaprakash-etal2019, Rodriguez-etal2020, Quintin-etal2021}.

The temporal variability of ULXs has been studied in several sources over the past few decades. However, interpretation of the physical origin of these variability properties remains challenging. For example, based on the observed variability in the power density spectra, \cite{Heil-etal2009} proposed that a sample of bright ULXs of comparable luminosity can be divided into two groups $-$ one showing very weak variability on the timescale of about $100$ s, whereas another exhibiting similar variability properties, observed in most of the black hole X-ray binaries (BH-XRBs) \citep{Belloni-etal2005, Nandi-etal2012, Athulya-etal2022, Aneesha-etal2024}. Intriguingly, the mechanism that suppresses the intrinsic variability in one group of ULXs remains under debate \citep{Heil-etal2009, Feng-etal2011}. Interestingly, the fractional variability amplitude ($f_{\rm var}$), indicative of the amount of variability, is found to be low ($\lesssim10\%$) in the hard-ultraluminous states (HUL) as compared to the soft-ultraluminous states (SUL) with $f_{\rm var}\sim 10-30\%$ \citep{Sutton-etal2013}. Further, several ULXs are also reported to exhibit Quasi-periodic Oscillations (QPOs) in the presence of flat-topped noise in the power density spectra \citep{Dewangan-etal2006, Strohmayer-etal2007, Feng-etal2010a, Rao-etal2010, Pasham-etal2015, Agrawal-etal2015, Atapin-etal2019, Majumder-etal2023}. In addition, it has been proposed that the observed QPOs are possibly generated due to the modulation of the Comptonizing corona, present at the inner accretion flow and that serves as a powerful tool to probe the accretion scenarios of the ULXs harboring black hole accretors \cite[]{Das-etal2021, Majumder-etal2023}.

Needless to mention that the spectral morphologies of ULXs are well studied in X-rays with dedicated missions like {\it XMM-Newton}, {\it Chandra} and {\it NuSTAR} over the years. Interestingly, the ULX spectra up to $\sim 10$ keV can be classified into two distinct types: one described by simple power-law and the other exhibiting a broad spectral curvature over the entire energy band \cite[]{Feng-etal2011}. Generally, a high power-law photon index of $\Gamma \gtrsim 3$ is seen for the sources of softer characteristics, whereas $\Gamma \sim 1$ is observed in a few harder sources. Notably, some of the hard power-law dominated ULXs show significant flux variability \citep{Feng-etal2009, Kaaret-etal2009, Soria-etal2009} similar to the hard state of BH-XRBs \citep{Belloni-etal2005,Remillard-etal2006}. Thus, the interpretation of power-law dominated spectral state in ULXs as the canonical hard state of BH-XRBs would imply a black hole of mass $\gtrsim 10^{3}$ $M_\odot$ \citep{Winter-etal2006, Feng-etal2011}. Meanwhile, a good number of high-quality observations from XMM-Newton could detect the soft excess in low energies ($< 2$ keV) and curvature around $\sim 3-10$ keV in ULX spectra (\citealt[and references therein]{Ghosh-etal2021}) unlike the BH-XRBs, showing a spectral turnover at relatively higher energies \citep{Xu-etal2019}. Further, it has been suggested that the high energy curvature could arise from the innermost hot accretion flow as a result of the interception of disc photons into the Comptonizing corona \citep{Middleton-etal2015, Mukherjee-etal2015, Jithesh-etal2022}. In contrast, the soft excess below $\sim 2$ keV is thought to be originated from the outflowing wind or entirely from the disc emission alone \citep{Poutanen-etal2007,Middleton-etal2015}.

The extensive monitoring with {\it XMM-Newton}, {\it Chandra}, {\it NuSTAR} and {\it Swift-XRT} enables the opportunity of studying the long-term spectro-temporal evolution of ULXs. So far, a handful of sources have been studied in quest of the long-term variation of ULX characteristics. For example, \cite{Gurpide-etal2021} reported the spectral evolution for a group of ULXs, revealing the origin of variability in these systems. In addition, the transition between several spectral states is conjectured from the long-term monitoring of a number of ULXs \citep{Yoshida-etal2010}. Further, the long-term monitoring of NGC $5408$ X$-1$ with {\it Swift-XRT} reveals dipping behavior in the light curve, possibly connected to the super-orbital phenomenon, similar to some of the BH-XRBs \citep{Grise-etal2013}.

In this work, we consider eight BH-ULXs, namely NGC $1313$ X$-1$, NGC $5408$ X$-1$, NGC $6946$ X$-1$, IC $342$ X$-1$, NGC $55$ ULX1, NGC $4395$ ULX1, NGC $5204$ X$-1$ and NGC $4190$ ULX1, having decade-long observations with {\it XMM-Newton}. The sources are primarily selected based on the publicly available {\it XMM-Newton} observations spanning over about a decade and the previous predictions of having black hole accretors at the central core of these objects. Out of this eight sources, a detailed spectro-temporal study is carried out for five sources (NGC $1313$ X$-1$, NGC $5408$ X$-1$, NGC $6946$ X$-1$, M $82$ X$-1$ and IC $342$ X$-1$) that show mHz QPO features \cite[hereafter Paper-I]{Majumder-etal2023}. Note that, the BH-ULX candidate NGC $247$ ULX$1$ was not studied in Paper-I due to the absence of any reported QPO detections for this source. It is worth mentioning that M$82$ X$-1$ is kept aside from the present work irrespective of having multiple observations because of the significant contamination from nearby objects in the {\it XMM-Newton} aperture (see \citealt[]{Majumder-etal2023} and references therein). Moreover, a preliminary analysis of the variability properties of NGC $247$ ULX$1$ suggests that the source exhibits a dipping behavior in its light curves (see \citealt[]{Alston-etal2021} also), accompanied by supersoft characteristics, which remain distinct from that of other sources. Therefore, we exclude NGC $247$ ULX$1$ in this study. Note that, the four sources, namely NGC $55$ ULX1, NGC $4395$ ULX1, NGC $5204$ X$-1$ and NGC $4190$ ULX1, have been relatively less studied and serve as ideal targets for detailed analysis. With the exception of the above source selection criteria, NGC $4190$ ULX1---often referred to as the `forgotten ULX' due to its intriguing yet relatively unexplored characteristics---is also considered in this work, despite having only three observations with {\it XMM-Newton}. Below, we present the brief characteristics of the four sources for which an in-depth analysis has been performed in this work.

\begin{itemize}
    \item NGC $55$ ULX$1$ is the brightest ULX in the spiral galaxy NGC $55$ at a distance of $1.94$ Mpc\footnote{\url{https://ned.ipac.caltech.edu}} with a peak luminosity of $\sim 4\times10^{39}$ erg $\rm s^{-1}$. Interestingly, in most of the previous studies, the source was found in the SUL state with disc-dominated spectra \citep{Jithesh-etal2022, Barra-etal2022}. 
	
    \item NGC $4395$ ULX$1$ is one of the ULXs located in the NGC $4395$ galaxy at a distance of $4.76$ Mpc with luminosity $\sim 3\times10^{39}$ erg $\rm s^{-1}$ \citep{Ghosh-etal2022}. Detailed spectral study with {\it XMM-Newton} and {\it Chandra} observations rules out the possibility of having an SUL state by confirming its steep power-law tail \citep{Earnshaw-etal2017}. 
	
    \item The source NGC $5204$ X$-1$ is located $\sim 15$ arcsec away from the centre of the host galaxy NGC $5204$ at a distance of $4.8$ Mpc. The evidence of significant outflows resulting in several emission features is observed in NGC $5204$ X$-1$ with {\it XMM-Newton} \citep{Kosec-etal2018}.
	
    \item The low surface brightness galaxy NGC $4190$ located at a distance of $3$ Mpc contains NGC $4190$ ULX$1$, a bright ULXs of luminosity $\sim (3-10)\times 10^{39}$ erg $\rm s^{-1}$ \citep{Ghosh-etal2021}. The mass of the source is predicted to be in the range of $10-30 M_\odot$, indicating the presence of a stellar-mass black hole in the system \citep{Ghosh-etal2021}. Recently, a super-Eddington slim disc scenario is also suggested for NGC $4190$ ULX$1$ \citep{Earnshaw-etal2024}.
\end{itemize}

In this work, we carry out a detailed long-term spectro-temporal analysis of the selected BH-ULXs using the archival {\it XMM-Newton} observations spanning over a decade. Towards this, we investigate the spectral characteristics of the sources by modeling the {\it XMM-Newton} spectra in $0.3-10$ keV energy range with the combination of physically motivated models, describing different emission mechanisms. Investigation of the spectral properties indicates the presence of both positive and negative correlations between the bolometric disc luminosity ($L_{\rm disc}$) and the disc temperature. The correlation properties are observed to be closely connected with the distinct spectral states, obtained from the long-term evolution of the sources. Finally, we find that the disc-corona-wind regulated accretion scenario described by the Keplerian and sub-Keplerian flow components is capable in explaining the spectro-temporal findings of the BH-ULXs.

The paper is organized as follows. In \S 2, we briefly mention the {\it XMM-Newton} observations of the sources and the standard data reduction procedure of {\it EPIC-PN} and {\it EPIC-MOS} instruments. The results of spectral and timing analyses are presented in \S 3. In \S 4, we infer the possible physical scenarios to delineate the findings and summarize the results in \S 5.

\section{Observation and Data reduction}


\begin{table*}
	\caption{Details of the {\it XMM-Newton} observations analyzed in this work of the four selected BH-ULXs. In the table, all the symbols have their usual meanings. The ticks in the remarks column indicate the cleaned and good-quality data available for the analysis. Here, $F_{\rm var}$ is not estimated for the observations affected by high particle background flares and the corresponding columns are filled with `$-$' symbol. Similarly, $\rm Total_{\rm rms}$ is not estimated when PDS is modelled only with a \texttt{constant} component and for such cases, respective column is filled using `$-$'. See the text for details.}
	
	\renewcommand{\arraystretch}{1.2}
	\resizebox{1.0\textwidth}{!}{%
		\begin{tabular}{l l c c c c c c c c c c c c c c c c c}
			\hline
			\multirow{2}{*}{Source} & \multirow{2}{*}{Mission} & \multirow{2}{*}{ObsID} & \multirow{2}{*}{Epoch} & \multirow{2}{*}{Date} & \multirow{2}{*}{MJD} & \multirow{2}{*}{Exposure} & \multirow{2}{*}{Rate (cts/s)} & \multirow{2}{*}{$F_{\rm var}$} & \multirow{2}{*}{$\rm Total_{\rm rms}$} & \multirow{2}{*}{Remarks} & \\ \\
			& & & & & & (ks) & ($0.3-10$ keV)& & ($\%$) &\\
			\hline
			
			
			NGC 55 ULX1 & {\it XMM-Newton} & 0028740201 & XMM1 & 2001-11-14 & 52227.63 & 30.4 & $1.80\pm0.23$ & $16.39\pm1.84$ & $24.11\pm2.88$ & $\checkmark$ \\
			
			&  & 0028740101 & XMM2 & 2001-11-15 & 52228.05 & $26.1$ & $1.94\pm0.33$ & $21.09\pm2.38$ & $22.61\pm0.92$ & $\checkmark$ \\
			
			&  & 0655050101 & XMM3 & 2010-05-24 & 55340.32 & 113.7 & $0.91\pm0.16$ & $10.16\pm6.15$ & $50.29\pm1.47$ & $\checkmark$ \\

			&  & 0824570101 & XMM4 & 2018-11-17 & 58439.62 & 108.9 & $0.66\pm0.14$ & $5.98\pm2.23$ & $45.36\pm1.73$ & $\checkmark$ \\

			&  & 0852610101 & XMM5 & 2019-11-27 &	58814.87 & 4.7 & $1.69\pm0.23$ & $10.02\pm1.32$ & $-$ & $\checkmark$ \\

			&  & 0852610201 & XMM6 & 2019-12-27 & 58844.52 & 4.8 & $1.72\pm0.24$ & $9.80\pm1.44$ & $-$ & $\checkmark$ \\

			&  & 0852610301 & XMM7 & 2020-05-11 & 58980.93 & 5.8 & $0.70\pm0.15$ & $9.34\pm3.94$ & $-$ & $\checkmark$ \\

			&  & 0852610401 & XMM8 & 2020-05-19 & 58988.40 & 4.8 & $1.66\pm0.34$ & $12.22\pm3.14$ & $-$ & $\checkmark$ \\

			&  & 0864810101 & XMM9 & 2020-05-24 & 58993.92 & 117.7 & $0.92\pm0.16$ & $16.42\pm6.29$ & $45.79\pm3.16$ & $\checkmark$ \\
			
			&  & 0883960101 & XMM10 & 2021-12-12 & 59560.31 & 130 & $1.14\pm0.11$ & $5.65\pm2.75$ & $48.21\pm13.77$ & $\checkmark$ \\

			\hline
			
			
			NGC 4395 ULX1 & {\it XMM-Newton} & 0112521901 & XMM1 & 2002-05-31 & 52425.03 & 13.9 & $0.27\pm0.09$ & $9.03\pm3.18$ &  $-$ & Poor statistics \\
			
			&  & 0112522001 & XMM2 & 2002-06-12 & 52437.76 & 17.1 & $-$ & $-$ &  $-$ & Obs. affected by particle flares \\
			
			&  & 0112522701 & XMM3 & 2003-01-03 & 52642.93 & 6.7 & $0.29\pm0.09$ & $11.23\pm8.24$ &  $-$ & Poor statistics \\
			
			&  & 0142830101 & XMM4 & 2003-11-30 & 52973.14 & 103.4 & $0.33\pm0.11$ & $4.11\pm1.19$ & $44.22 \pm 5.18$ & $\checkmark$ \\
			
			&  & 0744010101 & XMM5 & 2014-12-28 & 57019.42 & 53.5 & $-$ & $-$ &  $-$ & Source is offset in CCD frame \\
			
			&  & 0744010201 & XMM6 & 2014-12-30 & 57021.42 & 53 & $-$ & $-$ & $-$ & Source is offset in CCD frame \\

			&  & 0824610101 & XMM7 & 2018-12-13 & 58465.26 & 112.8 & $0.40\pm0.12$ & $16.83\pm6.87$ & $25.65 \pm 3.32$ & $\checkmark$ \\

			&  & 0824610201 & XMM8 & 2018-12-19 & 58471.24 & 113.6 & $0.47\pm0.11$ &  $81.17\pm4.59$ & $41.55 \pm 2.13$ & $\checkmark$ \\

			&  & 0824610301 & XMM9 & 2018-12-31 & 58483.22 & 110.5 & $0.38\pm0.27$ & $93.85\pm57.71$ & $45.18 \pm 6.28$ & $\checkmark$ \\

			&  & 0824610401 & XMM10 & 2019-01-02 & 58485.21 & 105.9 & $0.76\pm0.16$ & $25.67\pm6.64$ & $36.54 \pm 5.64$ & $\checkmark$ \\
			
		&  & 0913600101 & XMM11  & 2022-12-10 & 59923.34 & $33$ & $0.31\pm0.07$ &  $12.74\pm4.36$ &  $37.78\pm12.22$  & $\checkmark$ \\

            &  & 0913600501 & XMM12  & 	2022-12-14  & 59927.98 & $30$ & $0.31\pm0.16$ &  $10.65\pm5.24$ &  $35.58\pm6.53$  & $\checkmark$ \\

            &  & 0913600601 &  XMM13  & 2022-12-19 & 59932.32 & $38.8$ & $0.22\pm0.07$ &  $15.57\pm5.38$ &  $39.04\pm12.93$  & $\checkmark$ \\

            &  & 0913600701 & XMM14  & 2022-12-22 & 59935.65 & $28$ & $0.26\pm0.06$ &  $7\pm4$ &  $40.38\pm17.66$  & $\checkmark$ \\

            &  & 0913600801 & XMM15  & 2022-12-26 & 59939.86 & $28$ & $0.24\pm0.08$ &  $8.14\pm1.58$ &  $32.23\pm7.93$  & $\checkmark$ \\

            &  & 0913600901 & XMM16 & 2022-12-30 & 59943.59 & $28$ & $-$ &  $-$ &  $-$  & Obs. affected by particle flares \\

            &  & 0932391701 &  XMM17  & 2024-06-01 & 60462.76 & $43$ & $0.26\pm0.05$ &  $8.28\pm4.07$ &  $39.57\pm7.02$  & $\checkmark$ \\

            &  & 0932391801 & XMM18  & 2024-06-02 & 60463.69 & $36$ & $-$ &  $-$ &  $-$  & Obs. affected by particle flares \\

			\hline
			
			
			NGC 5204 X-1 & {\it XMM-Newton} & 0142770101 & XMM1 & 2003-01-06 & 	52645.05 & $16.9$ & $0.66\pm0.13$ & $4.90\pm1.38$ & $-$ & $\checkmark$ \\
			
			&  & 0142770301 & XMM2 & 2003-04-25 & 52754.57 & $\lesssim 1$ & $-$ & $-$ & $-$ & Low exposure data \\
			
			&  & 0150650301 & XMM3 & 2003-05-01 & 52760.18 & $2.9$ & $1.11\pm0.17$ & $4.16\pm3.13$ & $28.41\pm1.62$ & $\checkmark$ \\	
			
			&  & 0405690101 & XMM4 & 2006-11-15 & 54054.85 & 1.9 & $1.37\pm0.19$ & $\sim0.42$ & $-$ & $\checkmark$ \\
			
			&  & 0405690201 & XMM5 & 2006-11-19 & 54058.84 & 35.7 & $1.13\pm0.18$ & $5.11\pm1.23$ & $36.45\pm7.38$ & $\checkmark$ \\
			
			&  & 0405690501 & XMM6 & 2006-11-25 & 54064.82 & 15.9 & $0.83\pm0.14$ & $5.25\pm1.42$ & $46.96\pm3.73$ & $\checkmark$ \\
			
			&  & 0693851401 & XMM7 & 2013-04-21 & 56403.21 & $15$ & $0.66\pm0.13$ & $7.64\pm6.57$ & $-$ & $\checkmark$ \\
			
			&  & 0693850701 & XMM8 & 2013-04-29 & 56411.19 & 11.3 & $0.70\pm0.13$ & $5.01\pm1.22$ & $-$ & $\checkmark$ \\
			
			&  & 0741960101 & XMM9 & 2014-06-27 & 56835.95 & 21.3 & $0.64\pm0.13$ & $2.04\pm1.12$ & $-$ & $\checkmark$ \\

            &  & 0921360101 & XMM10 & 2023-05-18 & 60082.92 & 123 & $0.77\pm0.15$ & $2.74\pm1.12$ & $43.51\pm 7.88$ & $\checkmark$ \\

            &  & 0921360201 & XMM11 & 2023-11-10 & 60258.43 & 122 & $0.95\pm0.21$ & $4.82\pm1.78$ & $32.44\pm 8.72$ & $\checkmark$ \\
						
			\hline
			
			
			NGC 4190 ULX1 & {\it XMM-Newton} & 0654650101 & XMM1 & 2010-06-06 & 55353.51 & 21.1 & $-$ & $-$ & $-$ & Obs. affected by particle flares \\
			
			&  & 0654650201 & XMM2 & 2010-06-08 & 55355.47 & 9.8 & $1.05\pm0.17$ & $4.29\pm1.82$ & $-$ & $\checkmark$ \\
			
			&  & 0654650301 & XMM3 & 2010-11-25 & 55525.06 & 5.8 & $1.77\pm0.25$ & $6.53\pm1.12$ & $-$ & $\checkmark$ \\

			\hline

		\end{tabular}%
	}
	\label{table: Obs_details}

\end{table*}

\begin{figure}
    \begin{center}
    \includegraphics[width=\columnwidth]{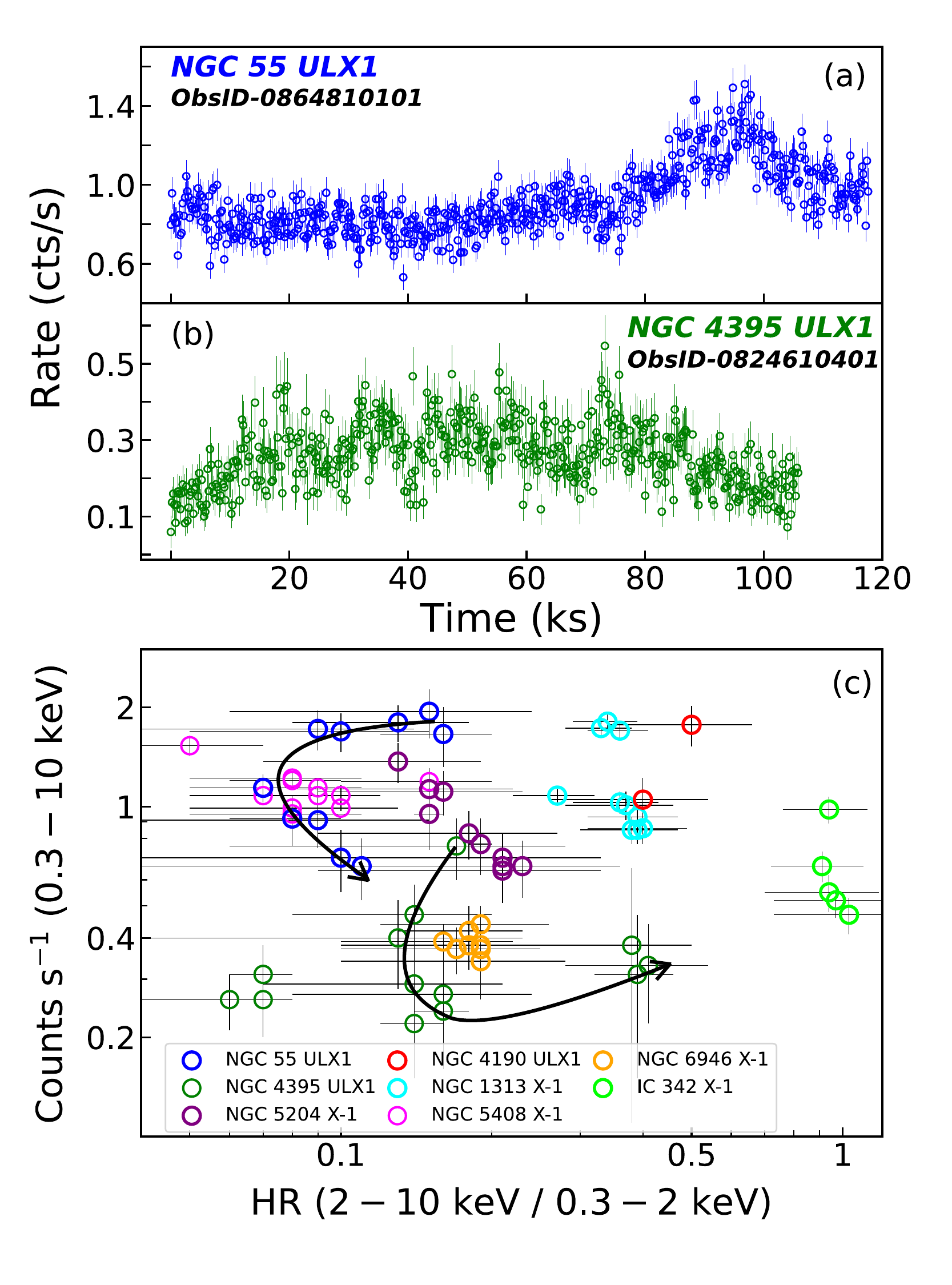}
    \end{center}
\caption{Panel (a)-(b): Background subtracted 200 s binned {\it EPIC-PN} light curves of NGC $55$ ULX$1$ and NGC $4395$ ULX$1$ in $0.3-10$ keV energy range. Panel (c): Hardness-intensity diagram (HID) of all eight BH-ULXs considered in this work. The open circles of different colors denote the HID of the respective sources. The results for NGC $6946$ X$-1$, NGC $1313$ X$-1$, IC $342$ X$-1$ and NGC $5408$ X$-1$ are adopted from Paper-I (\citealt{Majumder-etal2023}). In panel (c), the black curved arrows represent the `C-shaped' patterns observed in the HID of NGC $55$ ULX$1$ and NGC $4395$ ULX$1$, respectively. See the text for details.}
\label{fig:lcurve}
\end{figure}

We look into the \texttt{HEASARC} public data archive\footnote{\url{https://heasarc.gsfc.nasa.gov/db-perl/W3Browse/w3browse.pl}} for all the observations of NGC $55$ ULX1, NGC $4395$ ULX1, NGC $5204$ X$-1$ and NGC $4190$ ULX1 with {\it XMM-Newton}. We find that the sources are observed on several occasions by {\it XMM-Newton} and except NGC $4190$ ULX$1$, all the sources have at least ten or twenty years of monitoring. In Table \ref{table: Obs_details}, we tabulate the details of all the observations considered in this work. We find a total of thirty-one observations considering all the sources. However, we note that several observations suffer from various caveats (see Remarks on Table \ref{table: Obs_details}) and hence are excluded from the analysis.

The {\it XMM-Newton} data extraction of the remaining observations is carried out following the analysis threads\footnote{\url{https://www.cosmos.esa.int/web/xmm-newton/sas-threads}} provided by the instrument team. The data reduction software \texttt{SCIENCE ANALYSIS SYSTEM (SAS) V21.0.0}\footnote{\url{https://www.cosmos.esa.int/web/xmm-newton/sas}} is used to analyze the data. We run the tasks \texttt{epproc} and \texttt{emproc} to generate the event files for {\it EPIC-PN} and {\it EPIC-MOS} instruments, respectively. The high energy particle background flares are identified and clean event files are filtered out following \cite{Majumder-etal2023}. We adopt the event selection criteria as \texttt{PATTERN} $\leq 4$ and \texttt{PATTERN} $\leq 12$ with \texttt{FLAG}$=0$ while generating the scientific products from {\it EPIC-PN} and {\it EPIC-MOS} data, respectively. Following the previous studies, we select the source regions while extracting the light curves and spectra as $40$ arcsec, $25$ arcsec, $30$ arcsec and $40$ arcsec for NGC 55 ULX1 \citep{Jithesh-etal2022}, NGC 4395 ULX1 \citep{Ghosh-etal2022}, NGC 4190 ULX1 \citep{Ghosh-etal2021} and NGC 5204 X$-1$ \citep{Gurpide-etal2021}, respectively. Further, we select the same circular radii for the respective sources in a source-free area on the same {\it EPIC} chip to generate the background light curves and spectra. For spectral analysis, each spectrum is grouped with $25$ counts per spectral channel using the \texttt{specgroup}\footnote{\url{https://xmm-tools.cosmos.esa.int/external/sas/current/doc/specgroup/index.html}} tool of \texttt{SAS}.

\section{Results}

\subsection{Timing Analysis}

\label{s:timing}

\subsubsection{Variability and Hardness-Intensity Diagram (HID)}

\begin{figure}
    \begin{center}
    \includegraphics[scale=0.36]{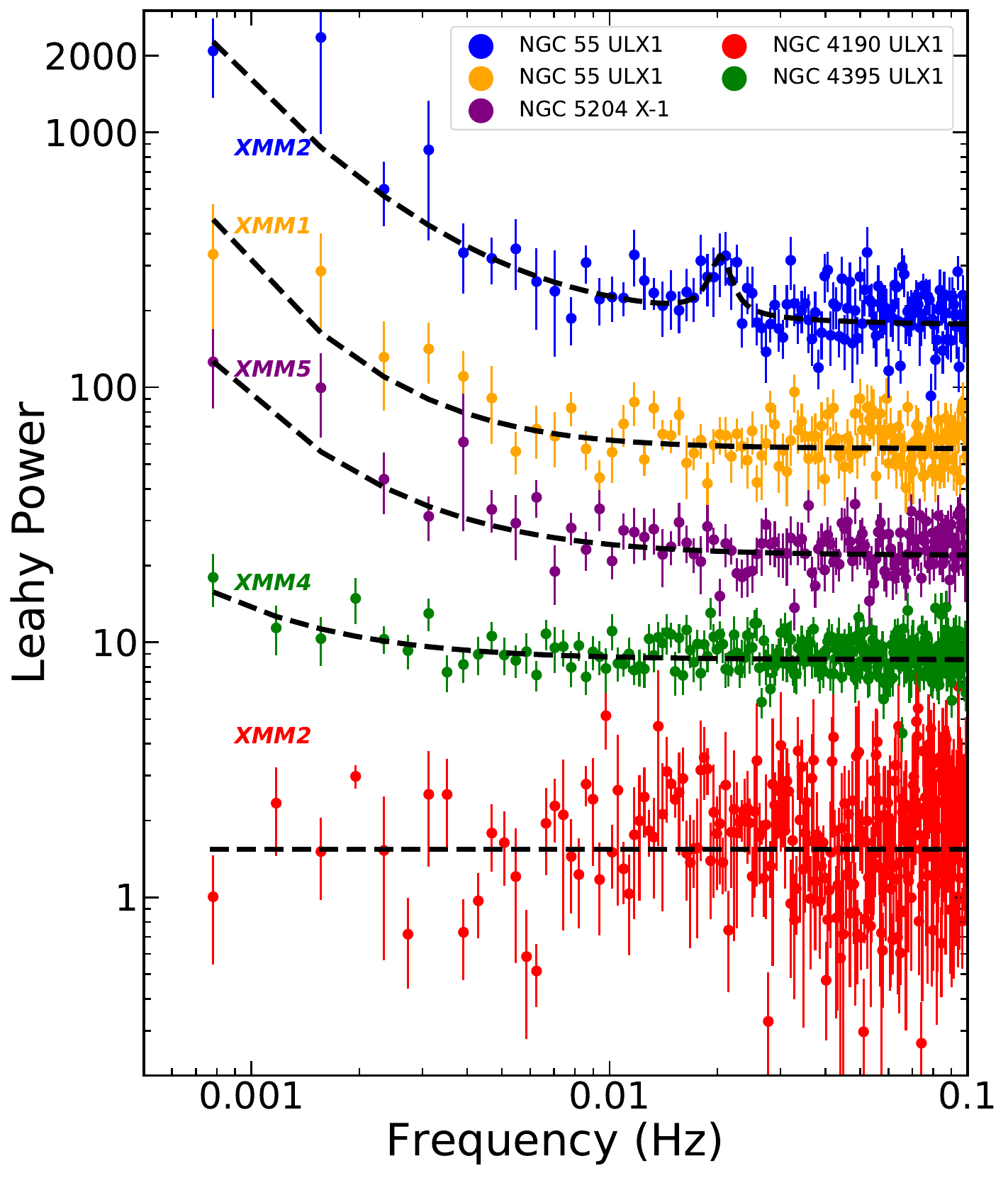}
    \end{center}
\caption{The PDS of NGC $55$ ULX$1$, NGC $5204$ X$-1$, NGC $4395$ ULX$1$ and NGC $4190$ ULX$1$ are depicted with different colors in respective Epochs. Each PDS is fitted with the combination of \texttt{constant}, \texttt{powerlaw}, and \texttt{Lorentzian} components as needed. The powers corresponding to the first four PDS presented from top to bottom are re-scaled with the constant multiplicative factors of $100$, $32$, $12$, and $4.5$, respectively, for better clarity. See the text for details.}
\label{fig:PDS}
\end{figure}

We study the variability properties of all the sources considered in this work. First, we generate background subtracted $200$ s binned {\it EPIC-PN} light curves in $0.3-10$ keV energy range. { Next, to deduce} the variability properties of the sources, we calculate the fractional rms variability amplitude, $F_{\rm var} = \frac{\sqrt{S^2-\overline{\sigma_{err}^2}}}{{\overline{x}}}$ following \cite{Vaughan-etal2003, Bhuvana-etal2021}. Here, $\overline{x}$, $S^{2}$ and $\overline{\sigma^{2}_{err}}$ are the mean count rate, variance, and associated error of the light curve. We find that the sources exhibit distinct variability properties with $F_{\rm var}$ varying in a wide range of $4.11-93.85\%$ including all the four sources and NGC $4395$ ULX1 remains the most variable source (see Table \ref{table: Obs_details}). Note that, a similar study on the BH-ULX variability indicates $F_{\rm var} \sim 1.42-27.28\%$ for the rest of the five sources (see Paper-I). In Fig. \ref{fig:lcurve}(a)-(b), we depict the background subtracted $200$ s binned light curves of NGC $55$ ULX$1$ and NGC $4395$ ULX$1$, respectively, for the representation of variability in BH-ULXs. It is noteworthy that for most of the sources marginal variability is observed and hence we refrain from presenting the light curves of the remaining sources in Fig. \ref{fig:lcurve}.

Further, we generate the hardness-intensity diagram (HID) by defining the hardness ratio as $HR=C_{1}/C_{2}$ with $C_{1}$ and $C_{2}$ being the background subtracted count rates in $2-10$ keV and $0.3-2$ keV energy ranges, respectively. In Fig. \ref{fig:lcurve}c, the HID of the respective sources are presented using different colored open circles. The data points in the HID of NGC $6946$ X$-1$, NGC $1313$ X$-1$, IC $342$ X$-1$ and NGC $5408$ X$-1$ are adopted from the results presented in Paper-I for comparison. We observe that each source shows a distinct pattern in the HID. In particular, NGC $55$ ULX1 and NGC $5408$ X$-1$ roughly show $HR \lesssim 0.15$ and IC $342$ X$-1$ exhibit a relatively harder nature with $HR \sim 1$, whereas the rest of the sources remain confined within $0.15\lesssim HR \lesssim 0.6$ in the HID. We observe that the HIDs of NGC $55$ ULX1 and NGC $4395$ ULX1 differ from the other sources, as they coarsely exhibit a `C-shaped' pattern (see Fig. \ref{fig:lcurve}c). Indeed, three of the most recent observations of NGC $4395$ ULX$1$ show fainter characteristics at higher energies ($\gtrsim 2$ keV) (see \S4.2) and remain as outlier ($HR\sim 0.07$) from the `C-shaped' pattern in the HID. Note that the intensity of these two sources varies significantly with $HR$, whereas the remaining sources do not show such variations. On the other hand, an apparent negative correlation is observed between $HR$ and the count rate of NGC $5204$ X$-1$, NGC $1313$ X$-1$ and IC $342$ X$-1$, respectively.

\subsubsection{Power Density Spectra}

The power spectral properties including QPO features are extensively studied for the five BH-ULXs namely NGC $1313$ X$-1$, NGC $5408$ X$-1$, M$82$ X$-1$, NGC $6946$ X$-1$ and IC $342$ X$-1$ in Paper-I. Following a similar approach, we investigate the power density spectrum (PDS) of the observations for the remaining four sources (NGC $55$ ULX$1$, NGC $5204$ X$-1$, NGC $4395$ ULX$1$ and NGC $4190$ ULX$1$) using {\it EPIC-PN} data in $0.3-10$ keV energy range. The PDS of the individual sources show a constant Poisson noise power within $1.64-1.96$ and a power-law distribution towards the low-frequency, which are modeled with a \texttt{constant} and \texttt{power-law} components, respectively. However, no significant variability is seen for a few observations of the sources (see Table \ref{table: Obs_details}), and a \texttt{constant} component is found to be sufficient to fit the PDS. In Fig. \ref{fig:PDS}, we depict the PDS of the sources in different colors for respective Epochs. Further, we estimate the total percentage rms amplitude of the PDS in the $0.001-0.1$ Hz frequency range for the individual sources and find it to vary within $22.61-50.29\%$.

\begin{itemize}
\item \textit{Possible detection of QPO:}  Interestingly, we notice that in one of the observation (XMM2) of NGC $55$ ULX1, significant excess in power remains near $\sim 0.02$ Hz after fitting the continuum with a \texttt{powerlaw} and \texttt{constant} components for a $\chi^{2}_{\rm red}$ of $167/125=1.34$. Hence, to fit the observed feature, we include one \texttt{Lorentzian} component near $\sim 0.02$ Hz. In general, each \texttt{Lorentzian} component is defined with three parameters namely the centroid frequency (LC), width (LW) and normalization (LN). With this, the resultant fit is obtained for an improved $\chi^{2}_{\rm red}$ of $156/122=1.28$ with the best-fitted \texttt{Lorentzian} centroid frequency of $20.43_{-1.67}^{+1.26}$ mHz. Next, we calculate the significance ($LN/err_{\rm neg}$, where, $err_{\rm neg}$ being the negative error in normalization), rms amplitude and Quality factor ($Q=LC/LW$) \citep{Belloni-etal2012, Sreehari-etal2019, Majumder-etal2022, Majumder-etal2023} of the fitted feature and obtained as $2.8\sigma$, $6.63\%$ and $6.68$, respectively. Based on the observed high Q-factor and strength of the feature ($rms\% \sim 6.63$), we interpret it as the possible detection of a QPO characteristics in NGC $55$ ULX$1$. For the remaining observations of the respective sources, power spectra remain featureless without any detection of QPO.
\end{itemize}

\subsection{Spectral Analysis}

\subsubsection{Spectral Modeling}

We model the {\it XMM-Newton} spectra in $0.3-10$ keV energy range of all the available observations for each source using \texttt{XSPEC V12.13.1} \citep{Arnaud-etal1996} in \texttt{HEASOFT V6.32.1}. The necessary background spectra, instrument response and ancillary files are also used while doing the spectral modeling. Both {\it EPIC-PN} and {\it EPIC-MOS} spectra are simultaneously fitted by including a \texttt{constant} component to adjust the calibration offset between the instruments. It is worth mentioning that the individual spectra of {\it EPIC-PN}, {\it EPIC-MOS1} and {\it EPIC-MOS2} are not combined for spectral fitting.

\begin{figure}
    \begin{center}
    \includegraphics[scale=0.3]{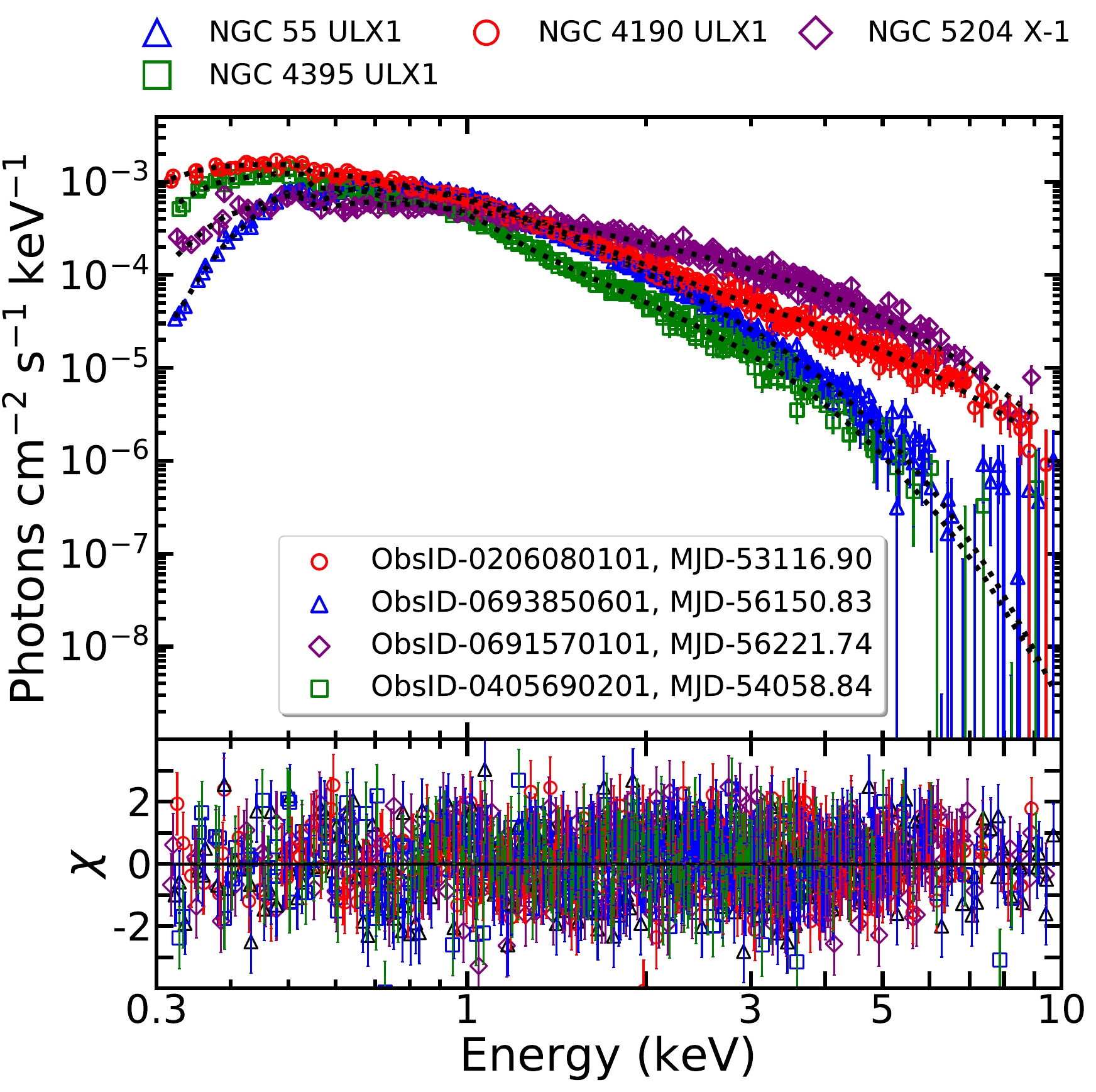}
    \end{center}
\caption{Top: Best fitted {\it EPIC-PN} and {\it EPIC-MOS} energy spectra of the four sources plotted simultaneously in $0.3-10$ keV energy range. Each spectrum is modeled using the model combination \texttt{constant$\times$Tbabs}$\times$(\texttt{diskbb}$+$\texttt{nthComp}). Bottom: Variation of residuals { (in units of $\sigma$)} obtained from the best fit of the individual spectra. See the text for details.}
\label{fig:Spectra}
\end{figure}

In the modeling, we consider the model combination, \texttt{constant$\times$Tbabs}$\times$(\texttt{diskbb}$+$\texttt{nthComp}), comprising of a thermal Comptonization component (\texttt{nthComp} in \texttt{XSPEC}) \citep{Zdziarski-etal1996} and a standard disc component (\texttt{diskbb} in \texttt{XSPEC}) \citep{Makishima-etal1986}. Here, \texttt{Tbabs} takes care of the inter-galactic absorption column and the absorption local to the source \citep{Wilms-etal2000}. It is worth mentioning that a similar model formalism is found to provide the best fit of the {\it XMM-Newton} spectra for other four sources (NGC $5408$ X$-1$, NGC $1313$ X$-1$, NGC $6946$ X$-1$ and IC $342$ X$-1$) as presented in \cite{Majumder-etal2023}. Hence, we proceed to generalize the model prescription for all the BH-ULXs considered in this work. While doing so, the seed photon temperature of \texttt{nthComp} component kept tied with the inner disc temperature of the \texttt{diskbb} component during the spectral modeling. With this, the above model prescription is seen to provide the best fit for the observations of all the sources under consideration. The best-fitted model parameters and the corresponding fit statistics are tabulated in Table \ref{tab:spectral1} for NGC $55$ ULX$1$, NGC $5204$ X$-1$, NGC $4395$ ULX$1$ and NGC $4190$ ULX$1$. We mention that the disc contribution is observed to be negligible and the Comptonization component remains adequate to delineate the spectra in a few observations (see Table \ref{tab:spectral1}). In contrast, we notice that for a few observations (XMM11$-$XMM17) of NGC $4395$ ULX$1$, the disc component alone remains adequate to fit the spectra. This is mostly because of the extremely faint nature of the source in higher energies ($\gtrsim 2$ keV) during these observations in which high-quality data remains up to $\lesssim 2$ keV only for spectral modeling. Additionally, we notice that the electron temperature remains unconstrained in most of the observations and hence, frozen to the best-fitted values. Notably, in a few observations (XMM1, XMM3, XMM4, XMM9) of NGC $55$ ULX$1$, an absorption \texttt{edge} is required to adjust the residuals near $\sim 1$ keV \citep{Jithesh-etal2022}. In contrast, a \texttt{Gaussian} line is used to fit the broad emission feature observed at $\sim 1$ keV in all the spectra of NGC $4395$ ULX$1$ \citep{Ghosh-etal2022}. The best-fitted energy spectra of NGC $55$ ULX$1$, NGC $4395$ ULX$1$, NGC $5204$ X$-1$ and NGC $4190$ ULX$1$ are shown in Fig. \ref{fig:Spectra} for the purpose of representation. Further, we compute the flux associated with different model components, optical depth, and Compton y-parameter following Paper-I. Also, the total and disc bolometric luminosities are estimated from the flux values, assuming the well-constrained source distances (see Table \ref{tab:spectral1}).

\subsubsection{Spectral Properties}


\begin{table*}
    \centering
    \renewcommand{\arraystretch}{1.4}
    \caption{Best-fitted and estimated spectral parameters obtained from the simultaneous modeling of {\it EPIN-PN} and {\it EPIN-MOS} spectra of the {\it XMM-Newton} observations of respective sources in $0.3-10$ keV energy range using the model combination \texttt{constant$\times$TBabs$\times$(diskbb + nthComp)}. All the quantities mentioned in the table have their usual meanings. The flux and luminosity values are presented in units of $\times10^{-12}$ erg $\rm cm^{-2}$ $\rm s^{-1}$ and $\times 10^{39}$ erg $\rm s^{-1}$, respectively.}
	
    \resizebox{2.0\columnwidth}{!}{%
		
    \begin{tabular}{l @{\hspace{0.3cm}} c @{\hspace{0.3cm}} c @{\hspace{0.3cm}} c @{\hspace{0.3cm}} c @{\hspace{0.3cm}} c @{\hspace{0.3cm}} c @{\hspace{0.3cm}} c @{\hspace{0.3cm}} c @{\hspace{0.3cm}} c @{\hspace{0.3cm}} c @{\hspace{0.3cm}} c @{\hspace{0.3cm}} c} \hline
    
			 Parameters & XMM1 & XMM2 & XMM3 & XMM4 & XMM5 & XMM6 & XMM7 & XMM8 & XMM9 &  XMM10 & XMM11 \\
			
			\hline
			&  &  &  &  &  NGC $55$ ULX$1$ &  &  &  & &  &\\
			&  &  &  &  &  (D $= 1.94$ Mpc) &  &  &  &  &  &\\
			\hline
			
			$n_{\rm H}$ ($10^{22}$ cm$^{-2}$) & $0.16^{+0.04}_{-0.03}$ & $0.33^{+0.05}_{-0.06}$ & $0.11^{+0.02}_{-0.02}$ & $0.15^{+0.03}_{-0.03}$ & $0.38^{+0.04}_{-0.04}$ & $0.41^{+0.12}_{-0.08}$ & $0.43^{+0.12}_{-0.13}$ & $0.40^{+0.11}_{-0.18}$ & $0.11^{+0.02}_{-0.02}$ & $0.11^{*}$ & $-$ \\
			
			$kT_{\rm in}$ (keV) & $-$ & $0.16^{+0.02}_{-0.03}$ & $-$ & $-$ & $0.15^{+0.03}_{-0.03}$ & $0.14^{+0.03}_{-0.02}$ & $0.14^{+0.01}_{-0.01}$ & $0.15^{+0.03}_{-0.04}$ & $-$  & $-$ & $-$ \\
			
			$\Gamma_{\rm nth}$ & $3.01^{+0.06}_{-0.05}$ & $2.65^{+0.02}_{-0.05}$ & $3.64^{+0.07}_{-0.07}$ & $3.42^{+0.05}_{-0.06}$ & $3.27^{+0.08}_{-0.07}$ & $3.31^{+0.15}_{-0.13}$ & $3.32^{+0.21}_{-0.30}$ & $3.31^{+0.26}_{-0.33}$ & $3.56^{+0.08}_{-0.07}$  & $3.48_{-0.04}^{+0.04}$ & $-$ \\
			
			$kT_{\rm e}$ (keV) & $1.62^{*}$ & $1.62^{*}$ & $1.62^{*}$ & $2^{*}$ & $1.89^{*}$ & $1.62^{*}$ & $2^{*}$ & $2^{*}$ & $1.62^{*}$  & $2^{*}$ & $-$ \\
			
			$\chi^2$/d.o.f & $338/283$ & $383/261$ & $470/319$ & $422/312$ & $207/169$ & $164/145$ & $152/107$ & $99/96$ & $456/337$  & $476/287$ & $-$ \\
			
			$F_{\rm disc}$ & $-$ & $2.88\pm0.19$ & $-$ & $-$ & $5.36\pm0.49$ & $9.03\pm0.41$ & $4.95\pm0.34$ & $7.40\pm0.34$ & $-$  & $-$ & $-$ \\
			
			$F_{\rm nth}$ & $6.21\pm0.14$ & $8.39\pm0.18$ & $2.89\pm0.07$ & $2.60\pm0.06$ & $9.32\pm0.21$ & $10.04\pm0.52$ & $3.63\pm0.17$ & $5.10\pm0.23$ & $2.88\pm0.19$  & $4.71\pm0.04$ & $-$ \\
			
			$F_{\rm bol}$ & $7.46\pm0.17$ & $16.04\pm0.36$ & $3.57\pm0.08$ & $3.39\pm0.08$ & $23.62\pm0.54$ & $33.78\pm0.75$ & $16.21\pm0.72$ & $21.76\pm1.47$ & $3.58\pm0.08$  & $4.75\pm0.03$ & $-$ \\
			
			$L_{\rm bol}$ & $3.36\pm0.08$ & $7.22\pm0.16$ & $1.60\pm0.04$ & $1.53\pm0.04$ & $10.64\pm0.24$ & $15.21\pm0.33$ & $7.30\pm0.32$ & $9.81\pm0.66$ & $1.61\pm0.06$  & $2.14\pm0.01$ & $-$ \\
			
			$L_{\rm disc}$ & $-$ & $2.44 \pm 0.09$ & $-$ & $-$ & $4.58 \pm 0.22$ & $8.10 \pm 0.18$ & $4.89 \pm 0.16$ & $6.51 \pm 0.15$ & $-$  & $-$ & $-$ \\

            $\tau$ & $8.32\pm 0.21$ & $9.72\pm0.23$ & $6.62\pm0.15$ & $6.30\pm0.13$ & $6.88\pm0.21$ & $7.42\pm0.41$ & $6.53\pm0.72$ & $6.55\pm0.79$ & $6.80\pm0.18$  &  $6.16\pm0.08$ & $-$  \\

            y-par & $0.88\pm0.04$ & $1.19\pm0.06$ & $0.55\pm0.03$ & $0.62\pm0.03$ & $0.70\pm0.04$ & $0.69\pm0.08$ & $0.67\pm0.15$ & $0.67\pm0.16$ & $0.58\pm0.03$  &  $0.59\pm0.02$ & $-$ \\
			
			\hline
			
			
			&  &  &  &  &  NGC $5204$ X$-1$ &  &  &  &  &  & \\
			
			&  &  &  &  &  (D $= 4.8$ Mpc) &  &  &  &   &  & \\
			
			
			\hline
			
			$n_{\rm H}$ ($10^{22}$ cm$^{-2}$) & $0.04^{+0.02}_{-0.02}$ & $-$ & $0.05^{+0.03}_{-0.02}$ & $0.06^{+0.01}_{-0.02}$ & $0.06^{+0.01}_{-0.01}$ & $0.04^{+0.01}_{-0.01}$ & $0.06^{+0.03}_{-0.03}$ & $0.04^{+0.01}_{-0.01}$ & $0.04^{+0.02}_{-0.02}$  & $0.06_{-0.01}^{+0.01}$ & $0.07_{-0.01}^{+0.01}$ \\
			
			$kT_{\rm in}$ (keV) & $0.26^{+0.04}_{-0.02}$ & $-$ & $0.28^{+0.06}_{-0.06}$ & $0.29^{+0.05}_{-0.03}$ & $0.28^{+0.03}_{-0.03}$ & $0.28^{+0.03}_{-0.03}$ & $0.24^{+0.04}_{-0.04}$ & $0.25^{+0.05}_{-0.04}$ & $0.25^{+0.04}_{-0.03}$  & $0.24_{-0.01}^{+0.01}$ & $0.25_{-0.02}^{+0.03}$ \\
			
			$\Gamma_{\rm nth}$ & $1.66^{+0.13}_{-0.06}$ & $-$ & $2.07^{+0.14}_{-0.16}$ & $2.16^{+0.29}_{-0.11}$ & $2.12^{+0.15}_{-0.16}$ & $1.85^{+0.11}_{-0.12}$ & $1.69^{+0.11}_{-0.09}$ & $1.77^{+0.13}_{-0.16}$ & $1.76^{+0.09}_{-0.11}$  & $1.79_{-0.05}^{+0.05}$ & $2.17_{-0.12}^{+0.12}$ \\
			
			$kT_{\rm e}$ (keV) & $1.62^{+0.38}_{-0.27}$ & $-$ & $2.5^{*}$ & $1.62^{*}$ & $2.07^{+1.04}_{-0.42}$ & $2.65^{*}$ & $1.73^{+0.49}_{-0.25}$ & $2.13^{*}$ & $1.98^{+0.61}_{-0.34}$  & $2.09_{-0.21}^{+0.29}$ & $2.49_{-0.43}^{+0.88}$ \\
			
			$\chi^2$/d.o.f & $236/225$ & $-$ & $167/164$ & $103/133$ & $327/328$ & $319/296$ & $223/212$ & $238/208$ & $268/247$  & $525/395$ & $393/332$ \\
			
			$F_{\rm disc}$ & $0.46\pm0.02$ & $-$ & $0.68\pm0.05$ & $0.95\pm0.02$ & $0.76\pm0.02$ & $0.57\pm0.01$ & $0.52\pm0.02$ & $0.42\pm0.02$ & $0.37\pm0.02$ & $0.85\pm0.02$ & $0.86\pm0.03$ \\
			
			$F_{\rm nth}$ & $1.29\pm0.03$ & $-$ & $1.95\pm0.09$ & $2.27\pm0.10$ & $2.10\pm0.05$ & $1.58\pm0.04$ & $1.35\pm0.06$ & $1.46\pm0.07$ & $1.34\pm0.03$ & $1.76\pm0.02$ & $2.41\pm0.02$ \\
			
			$F_{\rm bol}$ & $2.04\pm0.03$ & $-$ & $3.15\pm0.07$ & $3.76\pm0.09$ & $3.37\pm0.03$ & $2.64\pm0.06$ & $2.21\pm0.05$ & $2.24\pm0.05$ & $2.02\pm0.05$ & $2.61\pm0.03$ & $3.24\pm0.04$ \\
			
			$L_{\rm bol}$ & $5.62\pm0.08$ & $-$ & $8.69\pm0.02$ & $10.37\pm0.03$ & $9.29\pm0.08$ & $7.31\pm0.02$ & $6.09\pm0.14$ & $6.18\pm0.14$ & $5.57\pm0.16$ & $7.19\pm 0.05$ & $8.93\pm0.03$ \\
			
			$L_{\rm disc}$ & $1.71\pm0.06$ & $-$ & $2.48\pm0.14$ & $3.45\pm0.06$ & $2.78\pm0.06$ & $2.10\pm0.03$ & $1.98\pm0.06$ & $1.59\pm0.06$ & $1.71\pm0.05$ & $2.34\pm0.04$ & $2.37\pm0.05$ \\

        	$\tau$ & $18.37\pm3.26$ & $-$ & $10.47\pm1.11$ & $12.60\pm1.54$ & $11.27\pm2.42$ & $11.89\pm1.14$ & $17.23\pm3.17$ & $14.33\pm1.72$ & $15.04\pm2.35$  & $14.24\pm1.27$ & $3.31\pm0.28$ \\

                y-par & $4.27\pm1.07$ & $-$ & $2.14\pm0.46$ & $2.01\pm0.48$ & $2.05\pm0.42$ & $2.93\pm0.56$ & $4.01\pm0.83$ & $3.42\pm0.82$ & $3.51\pm0.67$  & $3.31\pm0.28$ & $1.89\pm0.28$ \\
   
			\hline

		\end{tabular}  
	} 
	\begin{list}{}{}
		\item $^{*}$Frozen to best-fitted value.
	\end{list}
	
	\label{tab:spectral1}
	
\end{table*}


\begin{table*}
	\centering
	\contcaption{}
	
	\renewcommand{\arraystretch}{1.4}
	
	\resizebox{2.0\columnwidth}{!}{%
		
		\begin{tabular}{l @{\hspace{0.3cm}} c @{\hspace{0.3cm}} c @{\hspace{0.3cm}} c @{\hspace{0.3cm}} c @{\hspace{0.3cm}} c @{\hspace{0.3cm}} c @{\hspace{0.3cm}} c @{\hspace{0.3cm}} c @{\hspace{0.3cm}} c @{\hspace{0.3cm}} c @{\hspace{0.3cm}} c @{\hspace{0.3cm}} c @{\hspace{0.3cm}} c @{\hspace{0.3cm}} c @{\hspace{0.3cm}} c @{\hspace{0.3cm}} c} 
			\hline
			
			
			Parameters & XMM2 & XMM3 & XMM4 & XMM7 & XMM8 & XMM9 & XMM10 & XMM11 & XMM12 &  XMM13 & XMM14 & XMM15 & XMM17  \\

			
			\hline
			&  &  & &  &  & NGC $4395$ ULX$1$ &  &  &  &  &    \\
			&  &  & & &  & (D $= 4.76$ Mpc) &  &  &  &  &    \\
			\hline
			
			$n_{\rm H}$ ($10^{22}$ cm$^{-2}$) & $-$ & $-$ & $0.10^{+0.03}_{-0.03}$ & $0.10^{+0.02}_{-0.02}$ & $0.10^{+0.02}_{-0.02}$ & $0.11^{+0.03}_{-0.02}$ & $0.11^{+0.03}_{-0.02}$ & $0.03_{-0.01}^{+0.01}$ & $0.03^{*}$ & $0.03^{*}$ & $0.04^{*}$ & $0.05^{*}$ & $0.05^{*}$ \\

            $kT_{\rm in}$ (keV) & $-$ & $-$ &  $0.15^{+0.02}_{-0.02}$ & $0.15^{+0.02}_{-0.02}$ & $0.16^{+0.03}_{-0.03}$ & $0.15^{+0.03}_{-0.03}$ & $-$ & $0.32_{-0.02}^{+0.02}$ & $0.31_{-0.01}^{+0.01}$ & $0.31_{-0.01}^{+0.01}$ & $0.31_{-0.01}^{+0.01}$ & $0.29_{-0.01}^{+0.01}$ & $0.31_{-0.01}^{+0.01}$ \\
   
			$\Gamma_{\rm nth}$ & $-$ & $-$ &  $3.34^{+0.17}_{-0.18}$ & $3.33^{+0.18}_{-0.16}$ & $3.20^{+0.09}_{-0.08}$ & $3.86^{+0.25}_{-0.29}$ & $3.05^{+0.05}_{-0.05}$ & $-$ & $-$ & $-$ & $-$ & $-$ & $-$  \\
			
			$kT_{\rm e}$ (keV) & $-$ & $-$ &  $2^{*}$ & $1.92^{*}$ & $2^{*}$ & $2^{*}$ & $1.62^{*}$ & $-$ & $-$ & $-$ & $-$ & $-$ & $-$  \\
			
			$\chi^2/d.o.f$ & $-$ & $-$ &  $179/155$ & $180/155$ & $259/212$ & $179/146$ & $313/237$ & $118/77$ & $96/86$ & $123/82$ & $141/87$ & $127/79$ & $131/108$\\
			
			$F_{\rm disc}$ & $-$ & $-$ &  $0.23\pm0.02$ & $0.24\pm0.02$ & $0.24\pm0.03$ & $0.27\pm0.03$ & $-$ & $0.55\pm0.01$ & $0.43\pm0.01$ & $0.41\pm0.02$ & $0.52\pm0.01$ & $0.49\pm0.01$ & $0.55\pm0.02$ \\
			
			$F_{\rm nth}$ & $-$ & $-$ &  $0.63\pm0.03$ & $0.62\pm0.02$ & $1.34\pm0.03$ & $0.74\pm0.03$ & $1.81\pm0.03$ & $-$ & $-$ & $-$ & $-$ & $-$ & $-$  \\
			
			$F_{\rm bol}$ & $-$ & $-$ & $1.37\pm0.03$ & $1.40\pm0.03$ & $2.37\pm0.03$ & $1.76\pm0.04$ & $2.71\pm0.04$ & $0.57\pm0.01$ & $0.45\pm0.01$ & $0.42\pm0.02$ & $0.54\pm0.01$ & $0.51\pm0.01$ & $0.57\pm0.02$ \\
			
			$L_{\rm bol}$ & $-$ & $-$ &  $3.71\pm0.08$ & $3.80\pm0.08$ & $6.43\pm0.08$ & $4.77\pm0.11$ & $7.35\pm0.08$ & $1.55\pm0.02$ & $1.22\pm0.01$ & $1.14\pm0.02$ & $1.46\pm0.02$ & $1.38\pm0.01$ & $1.55\pm0.02$ \\
			
			$L_{\rm disc}$ & $-$ & $-$ &  $1.27\pm0.05$ & $1.31\pm0.03$ & $1.28\pm0.04$ & $1.61\pm0.05$ & $-$ & $1.49\pm0.02$ & $1.17\pm0.01$ & $1.11\pm0.02$ & $1.41\pm0.02$ & $1.33\pm0.01$ & $1.49\pm0.02$  \\

                $\tau$ & $-$ & $-$ &  $6.48\pm0.41$ & $6.67\pm0.44$ & $6.83\pm0.23$ & $5.43\pm0.51$ & $8.13\pm0.16$ & $-$ & $-$ & $-$ & $-$ & $-$ & $-$ \\

                y-par & $-$ & $-$ &  $0.66\pm0.08$ & $0.67\pm0.09$ & $0.73\pm0.05$ & $0.46\pm0.08$ & $0.85\pm0.03$ & $-$ & $-$ & $-$ & $-$ & $-$ & $-$  \\

   			\hline
			&  &  & &  & & NGC $4190$ ULX$1$ &  &  &  &  &  &  \\
			&  &  & &  & & (D $= 3$ Mpc) &  &  &  &  &  &  \\
			\hline

                $n_{\rm H}$ ($10^{22}$ cm$^{-2}$) & $0.06^{+0.01}_{-0.01}$ & $0.15^{+0.06}_{-0.02}$ & $-$ & $-$ & $-$ & $-$ & $-$ & $-$ & $-$ & $-$ & $-$ & $-$ & $-$ \\
			
			$\Gamma_{\rm nth}$ & $2.09^{+0.19}_{-0.12}$ & $1.69^{+0.03}_{-0.03}$ & $-$ & $-$ & $-$ & $-$ & $-$ & $-$ & $-$ & $-$ & $-$ & $-$ & $-$ \\
			
			$kT_{\rm e}$ (keV) & $1.62^{*}$ & $1.67^{+0.11}_{-0.11}$ & $-$ & $-$ & $-$ & $-$ & $-$ & $-$ & $-$ & $-$ & $-$ & $-$ & $-$ \\
			
			$\chi^2$/d.o.f & $317/293$ & $335/337$ & $-$ & $-$ & $-$ & $-$ & $-$ & $-$ & $-$ & $-$ & $-$ & $-$ & $-$ \\
			
			$F_{\rm nth}$ & $3.31\pm0.06$ & $7.13\pm0.09$ & $-$ & $-$ & $-$ & $-$ & $-$ & $-$ & $-$ & $-$ & $-$ & $-$ & $-$ \\
			
			$F_{\rm bol}$ & $3.53\pm0.07$ & $8.17\pm0.12$ & $-$ & $-$ & $-$ & $-$ & $-$ & $-$ & $-$ & $-$ & $-$ & $-$ & $-$ \\
			
			$L_{\rm bol}$ & $3.80\pm0.07$ & $8.79\pm0.13$ & $-$ & $-$ & $-$ & $-$ & $-$ & $-$ & $-$ & $-$ & $-$ & $-$ & $-$ \\

                $\tau$ & $13.16\pm1.61$ & $17.56\pm0.79$ & $-$ & $-$ & $-$ & $-$ & $-$ & $-$ & $-$ & $-$ & $-$ & $-$ & $-$ \\

                y-par & $2.19\pm0.53$ & $4.03\pm0.23$ & $-$ & $-$ & $-$ & $-$ & $-$ & $-$ & $-$ & $-$ & $-$ & $-$ & $-$ \\

               \hline
			
		\end{tabular}  
	} 
    \begin{list}{}{}
		\item $^{*}$Frozen to best-fitted value.
    \end{list}
	
	
\end{table*}

In general, spectral features of the sources are well described by standard accretion disc and thermal Comptonization components, very similar to the spectral morphologies of a group of BH-ULXs presented in Paper-I. We find that the spectral modeling results the photon index ($\Gamma_{\rm nth}$) in the range of $1.66_{-0.06}^{+0.13}-3.86_{-0.29}^{+0.25}$ and the inner disc temperature ($kT_{\rm in}$) varies within $0.14_{-0.01}^{+0.01}-0.32_{-0.02}^{+0.02}$ keV for NGC $55$ ULX$1$, NGC $4395$ ULX$1$, NGC $5204$ X$-1$ and NGC $4190$ ULX$1$ (see Table \ref{tab:spectral1}). Note that, similar model formalism, used to fit the spectra of the remaining sources (NGC $1313$ X$-1$, NGC $5408$ X$-1$, NGC $6946$ X$-1$ and IC $342$ X$-1$), results in the electron temperature ($kT_{\rm e}$),  $\Gamma_{\rm nth}$ and $kT_{\rm in}$ in the range of $1.62-3.76$ keV, $1.48-2.65$ and $0.16-0.54$ keV, respectively (see Paper-I for details). Further, the optical depth and Compton y-parameter are found to be within $6 \lesssim \tau \lesssim 20$ and $0.46 \lesssim$ y-par $\lesssim 6.24$ considering all the sources (see Table \ref{tab:spectral1} and Paper-I). This suggests that the presence of relatively cool accretion disc and optically thick corona are the generic features of the BH-ULXs considered for this work \cite[]{Gladstone-etal2009}. 

\subsubsection{Spectral Correlation}

\begin{figure}
    \begin{center}
    \includegraphics[scale=0.4]{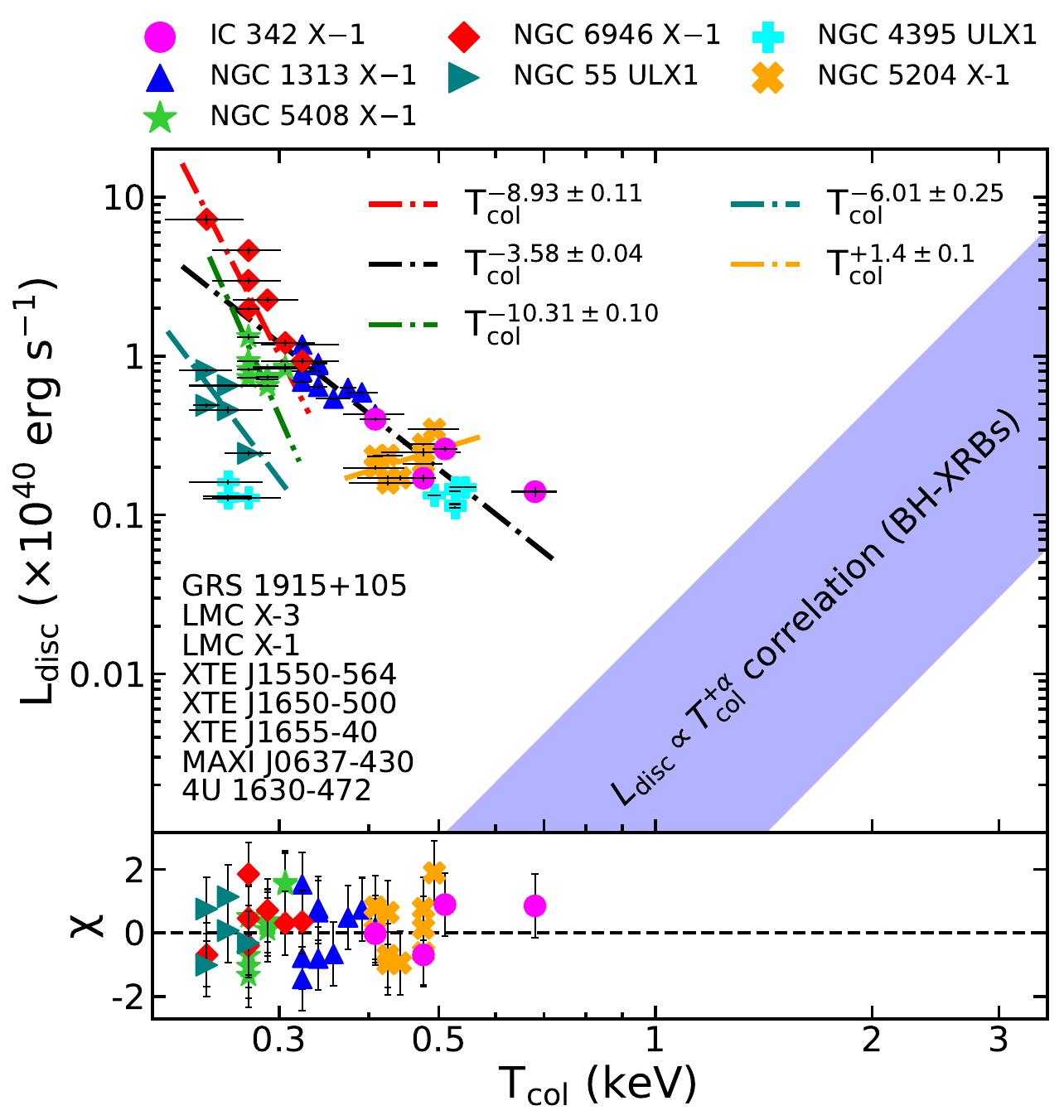}
    \end{center}
\caption{Correlation between the bolometric disc luminosity and color-corrected inner disc temperature of all BH-ULXs. The dash-dot lines of different colors represent the best fit in describing the $L_{\rm disc}-T_{\rm col}$ distribution of each source with the functional form mentioned in the figure. The blue-shaded region represents the correlation between the luminosity and disc temperature of the form $L_{\rm disc} \propto T_{\rm col}^{+\alpha}$ for eight BH-XRBs, adopted from \citealt{Gierlinski-etal2004,Abe-etal2005,Vierdayanti-etal2010,Baby-etal2021}. See the text for details.}
\label{fig:corr_plot}
\end{figure}

In this section, we attempt to deduce intrinsic correlations among the best-fitted and estimated spectral parameters for the respective sources. In doing so, we use the results obtained from the spectral analysis (see \S 3.2.1) of NGC $55$ ULX$1$, NGC $5204$ X$-1$ and NGC $4395$ ULX$1$, presented in Table \ref{tab:spectral1} along with the findings reported from the detailed spectral modeling of four BH-ULXs (NGC $1313$ X$-1$, NGC $5408$ X$-1$, NGC $6946$ X$-1$, IC $342$ X$-1$) in Paper-I (see Table-4). We investigate the correlation between the bolometric disc luminosity ($L_{\rm disc}$) and color-corrected disc temperature ($T_{\rm col}$) of the respective sources. Note that, the inner disc temperature ($T_{\rm in}$) is multiplied with the spectral hardening factor ($f_{\rm c}=1.7$) following \cite{Done-etal2008, Davis-etal2019} to obtain the $T_{\rm col}$. In Fig. \ref{fig:corr_plot}, we present the obtained results, where, the $L_{\rm disc}$ (in $\times 10^{40}$ erg $\rm s^{-1}$) is plotted as a function of $T_{\rm col}$ (in keV). 

We notice the trend of possible correlations of distinct characteristics between $L_{\rm disc}$ and $T_{\rm col}$ for all the sources. To deduce the firmness of the correlations, we estimate the Pearson correlation coefficient ($\rho$) of the distributions in $L_{\rm disc}-T_{\rm col}$ plane for the respective cases (see Paper-I). It is found that the source NGC $55$ ULX1 shows a negative correlation between $L_{\rm disc}$ and $T_{\rm col}$ with $\rho \sim -0.74$. However, the variation in $L_{\rm disc}$ is found to be marginal for NGC $4395$ ULX$1$ and the corresponding distribution in the $L_{\rm disc}-T_{\rm col}$ plane appears to form two distinct regimes of lower and higher temperatures (see Fig. \ref{fig:corr_plot}), indicating no evidence of correlation with $\rho \sim -0.11$ only. It may be noted that the higher disc temperatures ($\sim 0.3$ keV) are obtained for the observations in which only the disc component fits the entire spectra of narrow energy coverage ($\sim 0.3-2$ keV) in NGC $4395$ ULX$1$. Therefore, the two distinct regions of different disc temperatures (see Fig. \ref{fig:corr_plot}) perhaps resulted from the difference in model prescriptions for the respective cases. Contrarily, NGC $5204$ X$-1$ is seen to exhibit a clear positive correlation with $\rho \sim +0.7$. Moreover, a strong negative correlation is obtained with $\rho \sim -0.81$ for the combined results of NGC $1313$ X$-1$ and IC $342$ X$-1$, and $\rho \sim -0.89$ for NGC $6946$ X$-1$. Although, we find a weak anti-correlation with $\rho \sim -0.27$ in $L_{\rm disc}-T_{\rm col}$ plane of NGC $5408$ X$-1$. 

Further, we proceed to delineate the observed correlations of the sources with the empirical power-law profile $L_{\rm disc}\propto T_{\rm col}^{\alpha}$ \citep{Rybicki-etal1979} using the \texttt{curve\_fit}\footnote{\url{https://docs.scipy.org/doc/scipy/reference/generated/scipy.optimize.curve_fit.html}} module in \texttt{Python}. The individual fits result the power-law exponent ($\alpha$) as $-6.01\pm0.25$, $-8.93 \pm 0.11$, $-10.31 \pm 0.10$ and $1.4\pm0.1$ for NGC $55$ ULX1, NGC $6946$ X$-1$, NGC $5408$ X$-1$ and NGC $5204$ X$-1$, respectively. However, the combined fitting of the correlation distributions for IC $342$ X$-1$ and NGC $1313$ X$-1$ indicates $\alpha=-3.58\pm 0.04$. It may be noted that because of the marginal variation observed in $L_{\rm disc}$ and $T_{\rm col}$, we refrain from modeling the luminosity-temperature distribution for NGC $4395$ ULX$1$. In Fig. \ref{fig:corr_plot}, we present the best-fitted power-law functional form in the top panel for the respective sources and the residuals in the bottom panel, respectively. In addition, we show the positive correlation between the disc luminosity and temperature of the form $L_{\rm disc}\propto T_{\rm col}^{+\alpha}$, typically observed for BH-XRBs, with a blue shade in Fig. \ref{fig:corr_plot}. Note that the data associated with these correlation properties are adopted from \cite{Gierlinski-etal2004, Abe-etal2005, Vierdayanti-etal2010, Baby-etal2021} and the disc temperature values are color-corrected using the same hardening factor ($f_{\rm c}=1.7$), considered for the present work.

\begin{figure}
    \begin{center}
    \includegraphics[scale=0.42]{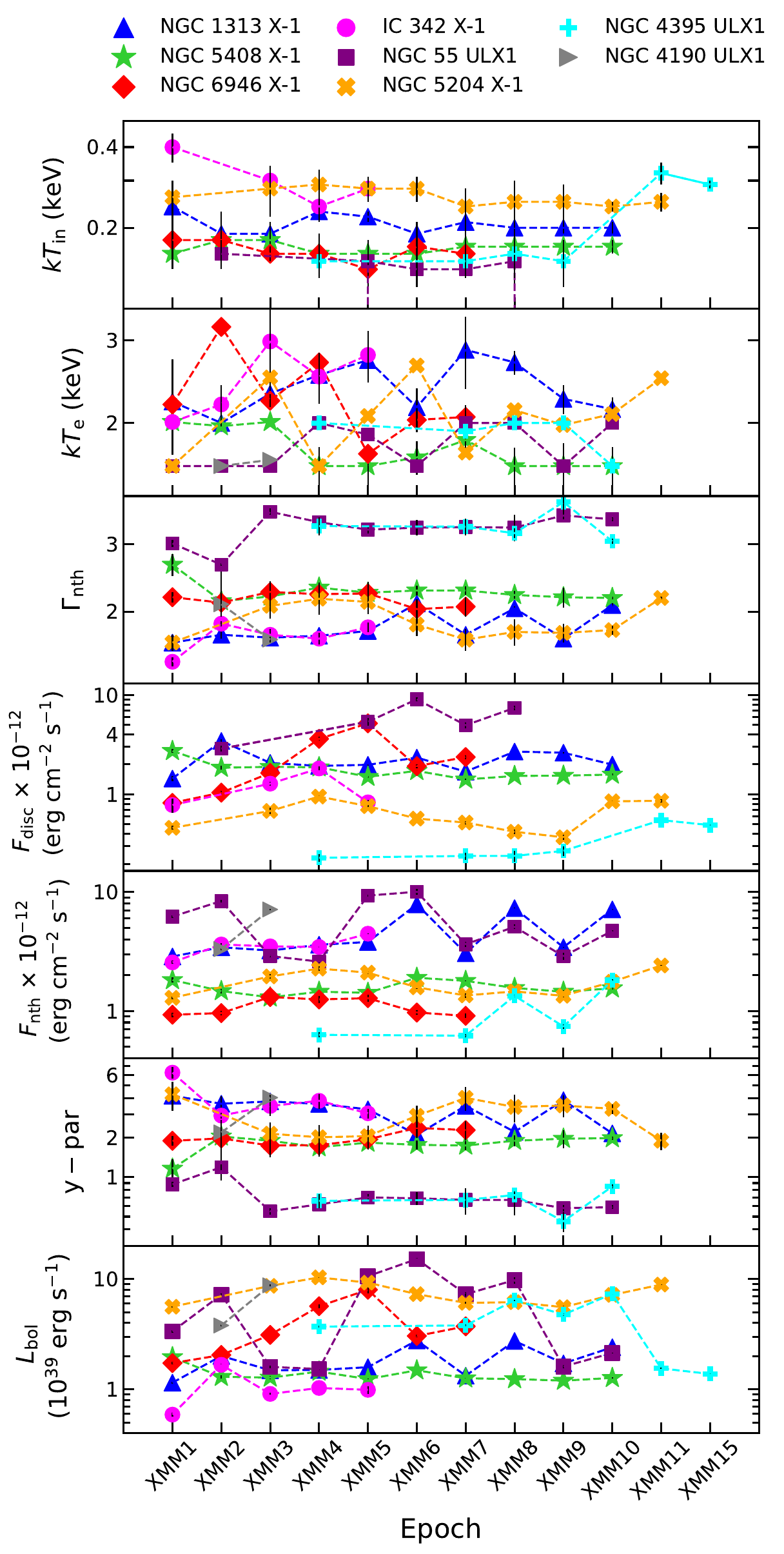}
    \end{center}
\caption{Evolution of the spectral parameters over the long-term monitoring are shown with distinct colored markers for the respective sources. The variation of inner disc temperature, electron temperature, photon index, disc flux, Comptonized flux, Compton y-parameter and bolometric luminosity over different Epochs are shown from top to bottom panels, respectively. See the text for details.}
\label{fig:spec_var}
\end{figure}

\subsubsection{Evolution of Spectral Properties}

We study the long-term spectral evolution of the eight BH-ULXs namely NGC $1313$ X$-1$, IC $342$ X$-1$, NGC $4395$ ULX$1$, NGC $5408$ X$-1$, NGC $55$ ULX$1$, NGC $4190$ ULX$1$, NGC $6946$ X$-1$ and NGC $5204$ X$-1$ using the {\it XMM-Newton} observations over a decade. In Fig. \ref{fig:spec_var}, the variation of inner disc temperature ($kT_{\rm in}$ in keV), electron temperature ($kT_{\rm e}$ in keV), photon index ($\Gamma_{\rm nth}$), disc flux ($F_{\rm disc}$ in erg $\rm cm^{-2}$ $\rm s^{-1}$), Comptonized flux ($F_{\rm nth}$ in erg $\rm cm^{-2}$$\rm s^{-1}$), Compton y-parameter (y-par) and bolometric luminosity ($L_{\rm bol}$ in erg $\rm s^{-1}$) are depicted from top to bottom panels, respectively, over different Epochs of observation. We mention that the best-fitted spectral parameters of NGC $1313$ X$-1$, IC $342$ X$-1$, NGC $5408$ X$-1$, and NGC $6946$ X$-1$ are adopted from the results presented in Paper-I (see Table-4 of the paper). However, for the rest of the sources, the best-fitted parameters are tabulated in Table \ref{tab:spectral1}.

We observe that the spectral properties of the BH-ULXs generally exhibit significant variation over the long-term evolution of the sources and manifest several spectral states. In particular, we find the dominance of the Comptonization flux over the disc contribution ($F_{\rm nth} \gtrsim F_{\rm disc}$) for NGC $5204$ X$-1$, IC $342$ X$-1$, NGC $4190$ ULX$1$ and NGC $1313$ X$-1$, respectively. As expected, the photon index is observed to be in the harder limit of $\Gamma_{\rm nth} \lesssim 2$, and the y-par, indicative of the amount of Comptonization, remains $\gtrsim 2$ for all the four sources (see Fig. \ref{fig:spec_var}). Further, we find a low disc temperature ($kT_{\rm in} \lesssim 0.3$ keV) for these sources except NGC $4190$ ULX$1$ in which the disc is absent and Comptonization contribution is noticeably high. In contrast, NGC $6946$ X$-1$ manifests disc dominated spectral characteristics with $F_{\rm disc} \gtrsim F_{\rm nth}$ and $\Gamma_{\rm nth} \gtrsim 2$ (see Fig. \ref{fig:spec_var}). Interestingly, comparable disc and Comptonized flux contributions are seen in the spectra of NGC $55$ ULX$1$ and NGC $5408$ X$-1$. However, the source NGC $55$ ULX$1$ replicates relatively softer nature with $\Gamma_{\rm nth} \gtrsim 3$ and y-par $\lesssim 1$, whereas NGC $5408$ X$-1$ remains in the intermediate regime ($F_{\rm disc} \approx F_{\rm nth}$, $\Gamma_{\rm nth} \lesssim 2$ and $1<$ y-par $\lesssim 2$). Finally, for NGC $4395$ ULX$1$, we observe a negligible disc contribution with y-par $\lesssim 1$ and high photon index ($\Gamma_{\rm nth} \gtrsim 3$). This possibly resulted from the appearance of broad emission feature at lower energy ($\sim 1$ keV), suggesting distinct and rather complex spectral characteristics of NGC $4395$ ULX$1$. Note that, the spectral parameters of NGC $4395$ ULX$1$ show marginal variations during later epochs (XMM11$-$XMM17) (see Fig. \ref{tab:spectral1}), hence we present only the results for XMM11 and XMM15 in Fig. \ref{fig:spec_var}. A significant variation in the luminosity is noticed over the entire observation period of more than a decade for most of the sources. Note that, a low electron temperature within $1$ keV $\lesssim kT_{\rm e} \lesssim 3$ keV is obtained for all the sources, indicating the presence of a relatively cool Compton corona, generally seen in ULXs.

\section{Discussion}

In this work, we study the long-term evolution of the spectro-temporal properties of eight BH-ULXs using {\it XMM-Newton} observations spanning over a decade or more. The detailed spectral analyses reveal the presence of significant correlations between the color-corrected inner disc temperature ($T_{\rm col}$) and the associated disc luminosity ($L_{\rm disc}$) of the sources. The long-term spectral evolution study infers different spectral states seems to be connected with the observed correlation properties of the respective sources.

We investigate the variability properties of four BH-ULXs, namely NGC $55$ ULX1, NGC $4395$ ULX1, NGC $5204$ X$-1$ and NGC $4190$ ULX1 in detail. For these sources, the fractional variability amplitude is found to vary as $4.11-93.85\%$. A similar study for the remaining sources ($i. e.$, NGC $1313$ X$-1$, NGC $5408$ X$-1$, NGC $6946$ X$-1$ and IC $342$ X$-1$) were already carried out in Paper-I \cite[]{Majumder-etal2023}, where the fractional variability is obtained as $1.42-27.28\%$. The HID presented in Fig. \ref{fig:lcurve}c indicates an apparent anti-correlation between the count rate and $HR$ for NGC $5408$ X$-1$, NGC $1313$ X$-1$, NGC $5204$ X$-1$ and IC $342$ X$-1$, respectively. Further, the power spectral study reveals the total percentage rms variability amplitude as $22-50\%$ in $0.001-0.1$ Hz frequency range for all BH-ULXs under consideration.

Furthermore, a detailed PDS investigation reveals the detection of a moderately significant ($\sigma \sim 2.8$, $rms\% \sim 6.63$) QPO-like feature at $\sim 20$ mHz in NGC $55$ ULX1 during Epoch XMM2. Interestingly, for this source, the PDS resembles the characteristics of a QPO (see \S\ref{s:timing}), although such feature is not observed within a day during its previous observation. It is worth mentioning that NGC $1313$ X$-1$, NGC $5408$ X$-1$, NGC $6946$ X$-1$, M$82$ X$-1$ and IC $342$ X$-1$ manifest prominent QPO features of frequency $\sim 8-667$ mHz over different observation Epochs of {\it XMM-Newton} \cite[and references therein]{Atapin-etal2019, Majumder-etal2023}.  

\subsection{Long-term Evolution of BH-ULXs}

The long-term evolution of spectral properties and the flux contributions from different spectral components confirm the presence of soft, intermediate, and hard spectral states in the BH-ULXs. In particular, we observe that NGC $5204$ X$-1$, IC $342$ X$-1$, NGC $4190$ ULX$1$ and NGC $1313$ X$-1$ remain within the group of ULXs, showing harder spectral features with dominant effects of thermal Comptonization (y-par $\gtrsim 2$) and relatively weak disc emission ($F_{\rm disc} \lesssim F_{\rm nth}$) (see Table \ref{tab:spectral1} and Fig. \ref{fig:spec_var}). Moreover, the flatter photon index ($\Gamma_{\rm nth} \lesssim 2$) possibly indicates the presence of a Comptonizing tail similar to BH-XRBs spectra in their hard/intermediate states \citep{Remillard-etal2006}. With this, we infer that for these BH-ULXs, an optically thick Comptonizing corona residing at the inner region of a low-temperature ($0.2 \lesssim kT_{\rm in} \lesssim 0.4$) disc appears to be the preferred accretion scenario with harder spectral characteristics.

In contrast, we find disc-dominated spectral morphologies in NGC $6946$ X$-1$, where the disc flux is roughly twice the Comptonized flux and $\Gamma_{\rm nth} \gtrsim 2$. Indeed, such classification of the sources based on the spectral flux contribution is similar to the soft-ultraluminous (SUL) and hard-ultraluminous (HUL) states \cite[]{Sutton-etal2013}. Notably, NGC $6946$ X$-1$ is observed in the disc-dominated state in the absence of a high-energy spectral break in most of the previous studies, and has been identified as a persistent ULX with softer characteristics \citep{Earnshaw-etal2019, Ghosh-etal2023}. Finally, the intermediate spectral nature ($F_{\rm disc} \approx F_{\rm nth}$, $\Gamma_{\rm nth} \lesssim 2$ and $1<$ y-par $\lesssim 2$) is observed for NGC $5408$ X$-1$ over the long-term monitoring which is identified as the intermediate-ultraluminous (IUL) state, whereas NGC $55$ ULX$1$ remains in the SUL state. We note that the presence of a `bright hard intermediate state' is reported for NGC $5408$ X$-1$ \citep{Caballero-etal2013} and NGC $55$ ULX$1$ is found to manifest a steep photon index ($\Gamma > 3$) in the previous studies \citep{Jithesh-etal2022} which are consistent with the present findings.

\subsection{Correlation between $L_{\rm disc}$ and $T_{\rm col}$ in BH-ULXs}

The variation of $L_{\rm disc}$ with $T_{\rm col}$ affirms the distinct correlation for all the BH-ULXs under consideration. In particular, all the sources, except NGC $5204$ X$-1$, show a negative correlation in the $L_{\rm disc}-T_{\rm col}$ plane, while a clear positive correlation is observed for NGC $5204$ X$-1$. The best-fitted correlations with the empirical power-law distribution $L_{\rm disc}\propto T_{\rm col}^{\alpha}$ yields the exponent as $\alpha=$ $-8.93$ (NGC $6946$ X$-1$), $-6.01$ (NGC $55$ ULX$1$), $-10.31$ (NGC $5408$ X$-1$), $+1.4$ (NGC $5204$ X$-1$) and $-3.58$ (combined IC $342$ X$-1$ and NGC $1313$ X$-1$), respectively (see Fig. \ref{fig:corr_plot}). It is noteworthy that NGC $4395$ ULX$1$ manifests marginal variation in the disc luminosity and temperature over different observations with $\rho \sim -0.11$ only, and hence we discard it from the present discussion.

We further note that a negative correlation between $L_{\rm disc}$ and $T_{\rm col}$ yielding a power-law exponent of $-3.5$ is reported for several ULXs \cite[]{Kajava-etal2009} that remains consistent with the present study. Interestingly, we observe an apparent connection between the correlation properties and the spectral states of the individual sources. For example, steeper ($\alpha < -4$) anti-correlations are seen for the sources in SUL (NGC $6946$ X$-1$ and NGC $55$ ULX$1$) and IUL (NGC $5408$ X$-1$) spectral states, respectively. However, a more flatter power-law exponent ($\alpha \sim -4$) is noticed for IC $342$ X$-1$ and NGC $1313$ X$-1$ in the HUL state. In addition, NGC $5204$ X$-1$ manifests a positive correlation in the HUL state.

Meanwhile, several galactic BH-XRBs, such as XTE J$1550-564$, XTE J$1650-500$, GRO J$1655-40$, GX $339-4$, XTE J$1859+226$ \cite[]{Gierlinski-etal2004}, and MAXI J$0637-430$ \cite[]{Baby-etal2021}, show clear positive correlations between $L_{\rm disc}$ and $T_{\rm col}$. This is also true for extra-galactic sources like LMC X$-1$ and LMC X$-3$ \cite[]{Gierlinski-etal2004}. These correlations align with the predictions of standard disc theory \citep{Shakura-etal1973}. Interestingly, depending on the luminosity, 4U $1630-472$ \citep{Abe-etal2005}, XTE J$1550-564$ \citep{Kubota-etal2004}, and GRS $1915+105$ \citep{Vierdayanti-etal2010} are observed to follow or deviate from the standard $L_{\rm disc} \propto T_{\rm col}^{4}$ correlation in different observations. In particular, when the luminosity is close to the Eddington limit, the sources are observed to exhibit a flatter correlation of the form $L_{\rm disc} \propto T_{\rm col}^{2}$ \citep{Watarai-etal2001}.

\subsection{Possible Dichotomy in $L_{\rm disc}\propto T_{\rm col}^{\alpha}$ Correlation}

In this study, we observe both positive and negative correlations in between $L_{\rm disc}$ and $T_{\rm col}$ for all BH-ULXs under consideration. Needless to mention that the appearance of `unusual' correlation including anti-correlations (`$-\alpha$') leads to the possible dichotomy between the present findings and the existing theoretical frameworks. The BH-XRBs are also seen to deviate from the standard picture on several occasions as pointed out in the preceding section. Nevertheless, several attempts have been made to explain this, although the origin of the dichotomy remains elusive. Depending on $\alpha$, the correlations are coarsely summarized in three different regimes for several BH-XRBs and BH-ULXs and briefly described as,
$$
L_{\rm disc} \propto T_{\rm col}^{\alpha};
\begin{aligned}
  \alpha &= 4, \qquad \qquad \text{(Standard disc)} \\   
  0 &\leq \alpha < 4, \qquad  \text{(Slim disc)} \\
  -10 &\leq \alpha < 0. \qquad \, \text{(Anomalous)} 
\end{aligned}
$$

\begin{itemize}
\item Standard disc scenario: Needless to mention that the luminosity associated with the disc emission is expected to follow $L_{\rm disc}\propto T_{\rm col}^{4}$ as predicted from the standard accretion disc prescription \citep{Shakura-etal1973}. In general, the Keplerian disc component of the accretion flow locally emits as a multi-color blackbody at different disc radii, resulting in the expected correlation \cite[]{Frank-etal2002}. 

\item Slim disc scenario: Several BH-XRBs including 4U $1630-472$, XTE J$1550-564$ and GRS $1915+105$ are seen to deviate from the standard disc and mostly follow a flatter correlation as $2 \lesssim \alpha < 4$ \cite[]{Kubota-etal2004, Abe-etal2005, Vierdayanti-etal2010}. It has also been suggested that different accretion scenarios exist for these BH-XRBs, which determine the nature of the correlations. In this study, we observe that NGC $5204$ X$-1$ is the only BH-ULX that manifests a positive correlation of the form $L_{\rm disc}\propto T_{\rm col}^{1.4}$, however this relation is not generally determined by the spectral states. Moreover, at relatively higher luminosity, close to the Eddington limit, the deviation from $\alpha \sim 4$ becomes more prominent \citep{Kubota-etal2004}. A geometrically thin ($H/r << 1$) and radiatively efficient standard Keplerian disc is characterized with moderately low accretion rate. However, the `slim disc' appears with relatively higher luminosity ($\gtrsim 0.3 L_{\rm Edd}$) yielding thicker disc geometry ($H/r \lesssim 1$), where the angular momentum remains sub-Keplerian \cite[]{Abramowicz-etal1988, Chakrabarti-etal1995a, Beloborodov-etal1998, Abramowicz-etal2013}. Interestingly, a more flatter correlation, $i.e.,~L_{\rm disc}\propto T_{\rm col}^{2}$, is predicted for such discs \citep{Watarai-etal2001}. Therefore, a possible explanation of the observed anomaly in the deviation of the power-law exponent could be the transition from the standard disc ($\alpha \sim 4$) to the `slim disc' scenario ($\alpha \sim 2$).

\item Anomalous scenario: Intriguingly, the negative correlations with $\alpha < 0$ observed in several BH-ULXs offer challenges, with only a few attempts made to explain these findings. \cite{King-etal2002b} proposed that the anti-correlation ($\alpha \sim -4$) is possibly resulted in due to the strong beamed emission in ULXs. Interestingly, the detection of coherent QPO features in BH-ULXs \cite[]{Atapin-etal2019, Majumder-etal2023} indicates the presence of axisymmetric geometrically thin accretion disc, which contradicts the presence of strong beaming \cite[]{Di_Matteo-etal1999, Strohmayer-etal2003}. Furthermore, \cite{Middleton-etal2019} reported that the Lense-Thirring precession of the inner accretion flow can attribute to the origin of the QPOs in ULXs. Meanwhile, \cite{Das-etal2021} argued that the aperiodic modulation of the inner hot `Comptonizing corona' successfully explains QPO features of BH-ULXs. All these findings clearly indicate the need for an alternative model prescription of accretion dynamics to explain the negative $\alpha$ values in the presence of QPO features.

\end{itemize}

\begin{table}
	\caption{Summary of the observed correlation properties between $L_{\rm disc}$ and $T_{\rm col}$ in different spectral states and the predicted inclinations of the BH-ULXs.}
	\renewcommand{\arraystretch}{1.2}
	\resizebox{1.0\columnwidth}{!}{%
		\begin{tabular}{l c c c c}
			\hline
			Source & Spectral & $L_{\rm disc}-T_{\rm col}$ & Predicted \\
                 & State$^\dagger$ & Correlation Exponent ($\alpha$) & Inclination \\
                \hline 

                NGC $5204$ X$-1$ & HUL & $1.4\pm0.1$ &  Low  \\
                IC $342$ X$-1$ & HUL & $-3.58\pm0.04$ &  Moderate \\
                NGC $1313$ X$-1$ & HUL & $-3.58\pm0.04$ &  Moderate  \\
                NGC $55$ ULX$1$ & IUL & $-6.01\pm0.25$ &  High  \\
                NGC $6946$ X$-1$ & SUL & $-8.93\pm0.11$ &  High  \\
                NGC $5408$ X$-1$ & IUL & $-10.31\pm0.10$ &  High  \\ 
                NGC $4190$ ULX$1$ & HUL & $-$ &  Low  \\
                M$82$ X$-1$ & $-$ & $-$ &  Low  \\
                NGC $4395$ ULX$1$ & HUL & $-$ &  Moderate  \\
			\hline	
		\end{tabular}  
	}
	\begin{list}{}{}
		\item HUL: Hard-ultraluminous state, SUL: Soft-ultraluminous state, IUL: Intermediate-ultraluminous state. $^\dagger$See \cite{Sutton-etal2013}
	\end{list}
 
	\label{table:corr}
\end{table}

\subsection{Alternative Physical Scenario}

\begin{figure}
    \begin{center}
    \includegraphics[scale=0.35]{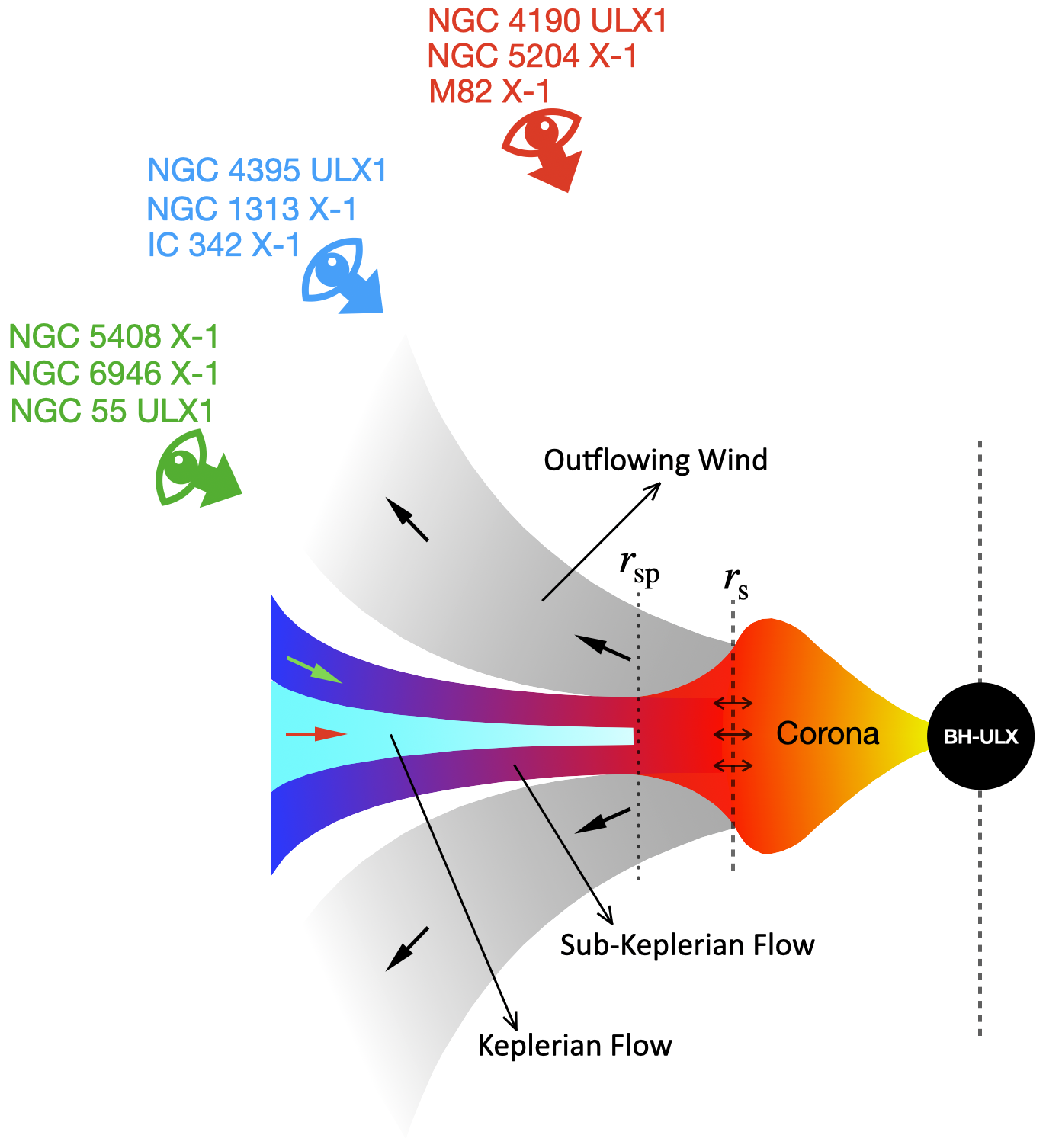}
    \end{center}
    \caption{Schematic representation of disc-corona-wind regulated two-component accretion scenario of BH-ULXs. Thick arrows in green, blue and red colors represent the viewing angles of the observer corresponding to high, moderate and low inclinations of sources, respectively. The red, green and black small arrows indicate the direction of the Keplerian flow, sub-Keplerian flow and winds. Here, $r_{\rm s}$ and $r_{\rm sp}$ denote the boundary of the corona and spherization radius. The region $r_{\rm sp} < r < r_{\rm s}$ refers the geometrically thick `slim disc'. The two-sided black arrows show the radial oscillation of the coronal region around $r_{\rm s}$. See the text for details.}
	\label{fig:cartoon}
\end{figure}

The limitations of the standard disc and slim disc models in explaining $L_{\rm disc}-T_{\rm col}$ correlations prompt us to explore alternative physically motivated accretion-ejection model prescriptions for BH-ULXs. Such a model prescription appears to be capable in  explaining the overall spectro-temporal properties, including $L_{\rm disc}-T_{\rm col}$ correlation and QPO features.

The accretion-ejection model formalism relies on the coexistence of Keplerian and sub-Keplerian flows \cite[]{Chakrabarti-etal1995b} and it explains the aspects of spectro-temporal variabilities of BH-XRBs \cite[and references therein]{Chakrabarti-etal2000, Chakrabarti-etal2009, Debnath-etal2010, Nandi-etal2012, Iyer-etal2015, Sreehari-etal2019}. Similarly, this model is expected to explain the spectro-temporal variabilities of BH-ULXs. Further, the non-steady modulation of the Comptonizing corona (with radius $r_{\rm s}$), formed by the sub-Keplerian component, leads to aperiodic variations in the emitted hard X-rays, which may give rise to QPO features \citep{Molteni-etal1996, Lee-etal2011, Das-etal2014, Garain-etal2014, Debnath-etal2024}. This conjecture effectively explains the mHz QPOs, which have become key observables for constraining the mass, spin, and accretion rate in several BH-ULXs \citep{Das-etal2021, Majumder-etal2023}. Moreover, by regulating the Keplerian and sub-Keplerian flow components and depending on system inclinations, different spectral states (HUL, SUL, and IUL) can be identified for BH-ULXs (see Table \ref{table:corr}). Furthermore, powerful winds are also expected to eject from (a) the boundary of hot coronal region in the inner accretion flow \citep{Chakrabarti-etal1999, Das-etal2001, Chattopadhyay-etal2002,Das-etal2014}, (b) standard thin Keplerian disc \citep{Proga-etal1998, Yang-etal2018, Raychaudhuri-etal2021} and (c) super-Eddington (super-critical) slim disc \citep{Poutanen-etal2007, Dotan-etal2011}, respectively.
Moreover, we mention that a model formalism in the framework of relativistic dissipative accretion flow around a rotating black hole has been developed to interpret the luminosity and mHz QPO characteristics of BH-ULXs \citep{Das-etal2021, Majumder-etal2023}. However, a more comprehensive modeling of the spectral energy distribution to explain the observed spectral characteristics of BH-ULXs is yet to be developed. We depict a schematic representation of the accretion-ejection configuration for the BH-ULXs in Fig. \ref{fig:cartoon}.

For super-critical accretion state of BH-ULXs, slim disc model seems relevant, where radiation pressure balances gravity at the spherization radius ($r_{\rm sp}$) resulting in a sub-Keplerian geometrically thick ($H/r \lesssim 1$) disc at $r_{\rm s} \leq r \leq r_{\rm sp}$ (see Fig. \ref{fig:cartoon} and \citealt[]{Fabrika-etal2021}). With this, it is apparent that the emission associated with the Keplerian disc ($r > r_{\rm sp}$) is consistent with the standard disc prediction ($L_{\rm disc} \propto T_{\rm col}^{4}$), whereas the interplay between the inner slim disc and the standard disc at larger radii likely leads to the deviation observed as $L_{\rm disc} \propto T_{\rm col}^{1.4}$ for NGC $5204$ X$-1$. As NGC $4190$ ULX$1$ and NGC $5204$ X$-1$ are observed in the HUL state throughout their long-term spectral evolution (see \S4.1), the inner Comptonizing corona, which is the source of non-thermal emission, at $r < r_{\rm s}$ is expected to be visible to the observer, suggesting a low inclination for these two sources (see Fig. \ref{fig:cartoon}). In addition, the absence of QPO features in NGC $4190$ ULX$1$ and NGC $5204$ X$-1$ possibly indicates that the radial modulation of the coronal region is either absent or very feeble to be detected. A similar accretion scenario seems to be relevant for M$82$ X$-1$ with the exception that oscillation of the corona is prominent enough to produce the observed mHz QPOs \cite[Paper-I,][]{Majumder-etal2023}. Indeed, the prediction of M82 X$-1$ inclination from the spectral analysis using {\it XMM-Newton} data could be uncertain because of high contamination. However, it may be noted that M82 X$-1$ exhibits coherent mHz QPO features on several occasions that are perhaps associated to the Comptonizing corona (see \citealt[]{Majumder-etal2023}). Further, the {\it Chandra} resolved spectra of the source prefers a slim disc scenario over the standard accretion disc \citep{Brightman-etal2016}. These evidences possibly suggest that M82 X$-1$ could be a low inclination system, where the inner Comptonizing corona coupled with a slim accretion disc is visible to the observer somewhat similar to the case of NGC $5204$ X$-1$.

In presence of winds, the emitted radiations from the disc interact substantially with the outflowing matter before reaching out to the observer. Due to this, the thermal emission is expected to be cooler with increasing luminosity for super-critical discs in the presence of strong outflowing winds \cite[]{Poutanen-etal2007}. Considering this, the bolometric luminosity ($L_{\rm bol}$) is estimated as,
\begin{equation}
    L_{\rm bol} \approx L_{\rm Edd} \left(1 + \frac{3}{5} \ln \dot{m}_{\rm 0} \right); \quad T_{\rm col} \approx 1.5 f_{\rm c} m^{-1/4} \dot{m}_{\rm 0}^{-1/2},
    \label{equ:1}
\end{equation}
where $T_{\rm col}$ denote the color-corrected temperature at the spherization radius and $\dot{m}_{\rm 0} ~(=\dot{M}_{\rm 0}/\dot{M}_{\rm Edd} \gg 1$) is the accretion rate. Here, $f_{\rm c}$, $L_{\rm Edd}$, and $m$ are the spectral hardening factor, Eddington luminosity, and mass of the central object scaled in solar mass, respectively. Note that equation (\ref{equ:1}) leads to an anti-correlation between $L_{\rm bol}$ and $T_{\rm col}$, with a steeper exponent ($\alpha$) \cite[]{Poutanen-etal2007}. Furthermore, the outflows from the region $r < r_{\rm sp}$ appear to remain optically thick, and a part of the energy generated within the disc is carried away by the winds. This process expected to reduce the disc temperature and modify the disc luminosity. 

The relatively flatter negative correlations ($\alpha \sim -3.58$) observed in the HUL state of IC $342$ X$-1$, NGC $1313$ X$-1$, and NGC $4395$ ULX$1$ suggest an accretion-ejection scenario, where the primary emission is significantly influenced by disc winds. However, the existence of the HUL state indicates that a part of the coronal emission is directly accessible to the observer, suggesting a moderate inclination of the sources (see Fig. \ref{fig:cartoon}, Table \ref{table:corr}). In contrast, for NGC $55$ ULX$1$, NGC $5408$ X$-1$, and NGC $6946$ X$-1$, the steeper negative correlations ($-10 \lesssim \alpha \lesssim -6$) in the SUL/IUL state likely indicate significant modification of the primary emission due to enhanced winds covering the line of sight. This suggests relatively higher inclinations for these sources (see Fig. \ref{fig:cartoon}, Table \ref{table:corr}). In both scenarios, it is plausible that the outflowing winds could carry the imprints of oscillations in the Comptonizing region, as predicted by \cite{Das-etal2014}, thereby providing a self-consistent explanation for the QPO features observed in IC $342$ X$-1$, NGC $1313$ X$-1$, NGC $5408$ X$-1$, and NGC $6946$ X$-1$. However, the irregular appearance of QPOs throughout the entire {\it XMM-Newton} monitoring of these sources \cite[see Paper-I,][]{Majumder-etal2023} suggests that either the necessary resonance condition for modulating the coronal region is not being satisfied, or the oscillation is weak enough to remain undetected.

Moreover, we speculate that the disc-corona-wind symbiosis for two-component super-critical accretion scenario seems to be realistic to explain the spectro-temporal features of BH-ULXs across different spectral states. We also indicate that radiative pressure-driven winds from the super-critical disc possibly govern the correlation between disc luminosity and temperature, and such correlation depends on the source inclination. This conjecture further strengthens the possibility of having a super-Eddington accretion scenario with massive stellar-mass to inter-mediate mass black hole accretors in NGC $5408$ X$-1$, NGC $6946$ X$-1$, IC $342$ X$-1$ and NGC $1313$ X$-1$, as predicted in \cite{Majumder-etal2023}.

\section{Conclusion}

In this study, we carry out a comprehensive analysis of the long-term evolution of eight BH-ULXs using {\it XMM-Newton} observations. Our findings provide compelling evidence for varied accretion scenarios that can explain the observed spectro-temporal variability and (anti-)correlations between disc luminosity ($L_{\rm disc}$) and inner disc temperature ($T_{\rm col}$). The key results from our investigation, along with their potential implications, are summarized below.

\begin{itemize}
    \item The positive correlation of $L_{\rm disc}\propto T_{\rm col}^{1.4}$ observed for NGC $5204$ X$-1$ suggests a transition from the standard geometrically thin disk ($H/r << 1$) to a slim disk ($H/r \lesssim 1$) prescription at relatively higher luminosities in the HUL state. This behavior is reminiscent of an accretion scenario where both Keplerian and sub-Keplerian flow components coexist.

    \item A steeper negative correlation with $-10 \lesssim \alpha \lesssim -6$ between $L_{\rm disc}$ and $T_{\rm col}$ is observed for sources NGC $6946$ X$-1$, NGC $55$ ULX$1$, NGC $5408$ X$-1$, which exhibit softer and intermediate spectral characteristics. In contrast, a relatively flatter anti-correlation with $\alpha \sim -3.58$ is seen for IC $342$ X$-1$ and NGC $1313$ X$-1$, which display harder spectral features.

    \item The diverse $L_{\rm disc}-T_{\rm col}$ correlations along with the overall spectro-temporal characteristics suggest a connection with the disc-corona-wind regulated two-component super-critical accretion scenario. Such scenario also potentially viable to explain the observed QPO features in BH-ULXs. In particular, we infer that the anti-correlations are possibly resulted due to the modification of disc emissions by the outflowing winds, which is also likely to be influenced by moderate to high inclinations of the respective BH-ULXs.
\end{itemize}

\section*{Acknowledgements}

Authors thank the anonymous reviewer for constructive comments and useful suggestions that help to improve the quality of the manuscript. SM, and SD thank the Department of Physics, IIT Guwahati, for providing the facilities to complete this work. AN thanks GH, SAG; DD, PDMSA, and Director, URSC for encouragement and continuous support to carry out this research. This publication uses the data from the {\it XMM-Newton} mission, archived at the HEASARC data center. The instrument team is thanked for processing and providing useful data and software for this analysis.

\section*{Data Availability}

Data used for this publication are currently available at the HEASARC browse website (\url{https://heasarc.gsfc.nasa.gov/db-perl/W3Browse/w3browse.pl}).

\input{ms.bbl}

\end{document}